\documentclass[11pt,a4paper, pdflatex]{article}
\usepackage{jcappub}

\usepackage{graphics}
\usepackage{xcolor}
\usepackage{float}
\usepackage{multirow}
\usepackage[normalem]{ulem}
\usepackage[T1]{fontenc}

\newcommand{\ssr}{Space Science Reviews}
\newcommand{\na}{New Astronomy}
\newcommand{\physrep}{Phys. Rept.}
\newcommand{\prd}{Physical Review D.}
\newcommand{\prl}{Physical Review Letters}
\newcommand{\jcap}{JCAP}
\newcommand{\apjl}{The Astrophysical Journal Letters}
\newcommand{\apjs}{The Astrophysical Journal Supplement Series}
\newcommand{\aap}{A\&A}%Astron. Astrophys.}
\newcommand{\apj}{ApJ} 
\newcommand{\aj}{The Astronomical Journal}

\newcommand\mnras{MNRAS}

\newcommand\apss{Astrophysics and Space Science}
\newcommand\apspr{Astrophysics and Space Physics Reviews}
\newcommand{\pasp}{Publications of the Astronomical Society of the Pacific}

\newcommand{\sz}{Sunyaev-Zel{'}dovich}

\newcommand{\sptpol}{SPTpol}
\newcommand{\spt}{SPT-3G}
\newcommand{\so}{SO}%-LAT}
\newcommand{\sofyst}{SO + FYST}%{SO-LAT+FYST}
\newcommand{\fyst}{FYST}

\newcommand{\cmbsfourwide}{S4-Wide}

\newcommand{\planck}{{\it Planck}}

\newcommand{\ukam}{$\mu$K-arcmin}
\newcommand{\bias}{b_{\rm lens}}
\newcommand{\mdplsimname}{\textsc{Agora}}
\newcommand{\msol}{M_\odot}

\newcommand{\srini}[1]{\textcolor{red}{[SR: #1]}}

\usepackage{array}
\newcolumntype{C}[1]{>{\centering\let\newline\\\arraybackslash\hspace{0pt}}m{#1}}
\newcommand{\mentiontablenotes}{} 
\ifdefined\mentiontablenotes
\usepackage[flushleft]{threeparttable} %for table notes
\fi

\newcommand{\refresponse}[1]{\textcolor{black}{#1}}
\newcommand{\addcomments}[1]{\textcolor{black}{#1}}

\title{A Foreground-Immune CMB-Cluster Lensing Estimator}

\author[a, b, 1]{Kevin Levy,\note{Corresponding author.}}
\author[c]{Srinivasan Raghunathan,}
\author[b]{Kaustuv Basu}

\affiliation[a]{School of Physics, University of Melbourne, Parkville, VIC 3010, Australia}
\affiliation[b]{Argelander-Institut f\"ur Astronomie, Universit\"at Bonn, D-53121 Germany}
\affiliation[c]{Center for Astrophysical Surveys, National Center for Supercomputing Applications, Urbana, IL 61801, USA}

\emailAdd{kevin.levy@student.unimelb.edu.au}
\emailAdd{srinirag@illinois.edu}
\emailAdd{kbasu@astro.uni-bonn.de}

\newcommand{\abstracttext}
{
Galaxy clusters induce a distinct dipole pattern in the cosmic microwave back-ground (CMB) through the effect of gravitational lensing. 
Extracting this lensing signal will enable us to constrain cluster masses, even for high redshift clusters ($z \gtrsim 1$) that are expected to be detected by future CMB surveys. 
However, cluster-correlated foreground signals, like the kinematic and thermal \sz{} (kSZ and tSZ) signals, present a challenge when extracting the lensing signal from CMB temperature data. 
While CMB polarization-based lensing reconstruction is one way to mitigate these foreground biases, the sensitivity from CMB temperature-based reconstruction is expected to be similar to or higher than polarization for future surveys.
In this work, we extend the cluster lensing estimator developed in \citet{raghunathan19c} to CMB temperature and test its robustness against systematic biases from foreground signals. 
We find that the kSZ signal only acts as an additional source of variance and provide a simple stacking-based approach to mitigate the bias from the tSZ signal.
Additionally, we study the bias induced due to uncertainties in the cluster positions and show that they can be easily mitigated. 
The estimated signal-to-noise ratio (SNR) of this estimator is comparable to other standard lensing estimators such as the maximum likelihood (MLE) and quadratic (QE) estimators. 
We predict the cluster mass uncertainties from CMB temperature data for current and future cluster samples to be: 6.6\% for \spt{} with 7,000 clusters, 4.1\% for \so{} and 3.9\% for \sofyst{} with 25,000 clusters, and 1.8\% for CMB-S4 with 100,000 clusters.
}

\defcitealias{raghunathan19c}{R19}
\defcitealias{madhavacheril18}{MH18}

\abstract{\abstracttext}

\begin{document}
\maketitle
\flushbottom

\section{Introduction}
Galaxy clusters are the largest gravitationally bound structures in the Universe. The evolution of the number density of galaxy clusters as a function of mass and redshift is a remarkable cosmological probe provided that their masses are accurately measured. 
In general, masses of galaxy clusters are inferred indirectly from observables like the \sz{} effect \citep{bleem15, plancksz15, huang20, bleem20, hilton20}, X-ray luminosity \citep{mantz10b}, and cluster richness, which is a measure of the number of galaxies inside a cluster \citep{rozo10, rykoff14}. 
The process of converting these observable quantities into mass involves certain assumptions about complex astrophysics that are not yet well understood and hence prone to systematic errors \citep{linden14}. 

Gravitational lensing, which fully traces the matter distribution in clusters, has proven to be a powerful tool to obtain unbiased cluster masses \citep{bartelmann96}. 
Significant efforts have also been undertaken to use lensing to calibrate the observable-mass relations of the quantities listed above in order to extract cosmological constraints from clusters %\citep[recently, ][]{zubeldia19, bocquet19, , costanzi21}. 
\citep{zubeldia19, bocquet19, to21, costanzi21}. 
Given that the recently launched and future surveys are forecasted to increase the sample size of galaxy clusters by more than two orders of magnitude, it is crucial to eliminate the systematic errors in their mass measurements. 
These surveys include cosmic microwave background (CMB) experiments such as \spt{} \citep{benson14, bender18, sobrin22}, AdvACTPol \citep{henderson16}, 
Simons Observatory (\so) \citep{so19}, Fred Young Submillimeter Telescope (\fyst) \citep{choi20, aravena23}, formerly called Cerro Chajnantor Atacama Telescope (CCAT-prime), and CMB-S4 (\cmbsfourwide) \citep{cmbs4-sb1, abazajian19}; optical and near-infrared surveys like Euclid \citep{laureijs11} and Vera C. Rubin Observatory \citep{lsst09}; and X-ray surveys such as eRosita \citep{predehl10}.

Measurements of gravitational lensing require a background light source behind the cluster. 
Optical weak lensing probes the total mass of a cluster through a statistical analysis of the image distortions induced on background galaxies by the foreground cluster \citep{hoekstra13}. 
However, since the signal-to-noise ratio (SNR) of the background galaxies that can be used for weak lensing measurements decreases with redshift, it is challenging to measure masses of high-redshift clusters ($z \gtrsim 1$) that are expected to be detected by future surveys.
%%%Furthermore, uncertainties in the photometric redshift estimation of the background galaxies, which increase with redshift, lead to additional errors in the mass estimation \citep{koehlinger15}. 

An effective alternative is to use the CMB as light source \citep{seljak00b, dodelson04}. 
The statistical properties of the CMB are well understood and since it originates behind all clusters in the Universe at $z \approx 1100$,  it constitutes a powerful tool to constrain masses of clusters with $z \gtrsim 1$ \citep{lewis06a}. 
Several estimators have been developed to extract cluster lensing signatures from CMB maps. 
Among these are matched filtering techniques \citep{seljak00b, holder04, vale04, horowitz19}; 
%maximum likelihood estimators (MLE), which fit cluster-lensed CMB templates to the observed maps \citep{dodelson04, lewis06a, baxter15, raghunathan17a}; quadratic estimators (QE), making use of the lensing induced correlation between the large-scale unlensed CMB gradient and the small-scale lensing signal \citep{maturi05, hu07, yoo08, yoo10, melin15}; background CMB gradient inversion techniques \citep{horowitz19}; 
maximum likelihood estimators (MLE) \citep{dodelson04, lewis06a, baxter15, raghunathan17a}; 
quadratic estimators (QE) \citep{maturi05, hu07, yoo08, yoo10, melin15};
and deep learning approaches \citep{gupta20}. 
%\addcomments{Include Horowitz19 with matched filtering technqiues since it says: \emph{we show that a simple gradient inversion matched filtering approach, as proposed by Seljak and Zaldarriaga(2000), ...}?}
More recently, \citet[hereafter \citetalias{raghunathan19c}]{raghunathan19c} developed a real-space estimator which fits lensing dipole templates to observed dipoles. 
\addcomments{This approach is analogous to the gradient inversion matched filter proposed earlier by \citet{seljak00b, horowitz19}.}

One limitation of using the CMB as background source is that the SNR of the CMB-cluster lensing signal for a single cluster is small (only $\sim$ 10 $\mu$K even for a cluster mass $\sim10^{15}\ \msol$). Therefore, a large number of clusters have to be stacked to achieve a reasonable SNR.
Following the first detection of the signal in 2015 by ACT \citep{madhavacheril15}, SPT \citep{baxter15}, and \planck{} \citep{plancksz15}, several groups have detected the signal using CMB temperature data \citep{geach17, raghunathan17b, baxter18, raghunathan19a, geach19, madhavacheril20}.
%%The detection was obtained by stacking the lensing signal extracted using CMB temperature maps at the location of many objects such as clusters \citep{baxter15, plancksz15, geach17, baxter18, raghunathan19a, madhavacheril20}, galaxy groups \citep{madhavacheril15, raghunathan17b}, and quasars \citep{geach19}. \sout{Recently,} \citetalias{raghunathan19c} \sout{reported the first detection of the cluster lensing signal using solely CMB polarization maps, which were obtained from the \sptpol{} survey.}
The cluster lensing signal can also be observed in CMB polarization. However, since the CMB gradient in polarization is roughly an order of magnitude lower than in temperature, and the strength of the lensing signal is directly proportional to the amplitude of the background gradient, the lensing SNR in polarization is much lower compared to temperature.
\citetalias{raghunathan19c} reported the first and only detection of the cluster lensing signal using solely CMB polarization maps, which were obtained from the \sptpol{} survey.
%\srini{Trim this a little.}

%While galactic foregrounds can be neglected in CMB-cluster lensing, extragalactic foregrounds have to be taken into account, such as emission from radio galaxies, emission from dusty star-forming galaxies commonly referred to as cosmic infrared background (CIB), and the thermal and kinematic Sunyaev-Zel’dovich (tSZ and kSZ) effects \citep{sunyaev70b, sunyaev80b}. All these components will add an effective noise floor in the temperature maps. Additionally, the clusters own tSZ and kSZ effect can bias the mass result and therefore have to be removed properly.

%\sout{Taking into account the effects of foreground signals is important when extracting the CMB-cluster lensing signal.}
While the lensing SNR is higher in temperature, it is important to take into account the effects of foreground signals in CMB temperature maps.
In particular, CMB temperature-based lensing reconstruction is contaminated by astrophysical foreground signals that are correlated with the cluster under study. 
These include the kinematic and thermal Sunyaev-Zel’dovich (kSZ and tSZ) signals \citep{sunyaev70b, sunyaev80b}, and emission from galaxies associated with the cluster. 
%%%Note that these foreground signals are largely unpolarized \citep{datta19, gupta19}, and hence CMB polarization can provide robust lensing reconstruction \citep{raghunathan17a}.
%\sout{However, the lensing SNR, which depends on the magnitude of the background CMB gradient at the cluster location, is lower for polarization compared to temperature. For that reason, CMB temperature-based lensing measurements will continue to dominate the SNR until the effective CMB map noise levels drop below $\Delta_{T} \sim$ 2 \ukam{}} \citep[see right panel of Fig. 2 from][]{raghunathan17a}.
%%%However, as discussed above, the lensing SNR is lower in polarization, and as a result, CMB temperature-based lensing measurements will continue to dominate the SNR until the effective CMB map noise levels drop below $\Delta_{T} \sim$ 2 \ukam{} \citep[see right panel of Fig. 2 from][]{raghunathan17a}.
Hence, it is important to develop strategies to mitigate the biases from foreground signals when using CMB temperature data. 

\citet{madhavacheril18} modified the QE to eliminate the bias from the cluster's tSZ signal. 
This modification involves estimating the large-scale background CMB gradient from a tSZ-free map. 
Since the QE uses lensing-induced correlations between large- and small-scales, eliminating the tSZ signal in the large-scale leg of the QE fully eliminates the bias. 
Such a tSZ-free map can be constructed using data from multiple frequency bands by making use of the frequency dependence of the tSZ signal. 
On the other hand, a bias from the kSZ signal, which has the same frequency dependence as the CMB, cannot be eliminated using the same method. 
\citet{raghunathan19b} made further modifications to the QE by estimating the large-scale gradient from an inpainted map that is free from all cluster-correlated foreground signals. 

In this work, we analyze and extend the lensing estimator constructed in \citetalias{raghunathan19c} such that it can be applied to CMB temperature data. 
We show that the estimator can be trivially modified to remove biases from cluster-correlated foreground signals and compare its performance with that of standard lensing estimators.  
%\sout{Additionally, we forecast the expected cluster mass constraints from upcoming CMB experiments such as \so, \sofyst, and \cmbsfourwide, and compare them to the constraints obtained for \spt, which is already collecting data.}
%\srini{You are not really comparing them to SPT-3G but just forecasting the values for each experiment, right?}
%\maybe{Using realistic simulations that include foregrounds}
Additionally, we forecast the expected cluster mass constraints for \spt{} and upcoming CMB experiments such as \so, \sofyst, and \cmbsfourwide.
%Finally, we show forecasts for upcoming CMB experiments such as CMB-S4 Wide (\cmbsfour), Simons Observatory (SO), and CCAT-prime (\ccatprime).

This paper is structured as follows: we give a brief overview of CMB cluster-lensing in \S\ref{sec:theory}; the simulations to which the estimator is applied are described in \S\ref{sec:simulations}; the lensing estimator itself is described in \S\ref{sec:estimator}; we validate the lensing pipeline in \S\ref{sec:validation}, compare the estimator with standard CMB-cluster lensing estimators like the MLE and QE in \S\ref{sec:comparison}, analyze the influence of uncertainties in the cluster positions in \S\ref{sec:positions}, and discuss foreground systematics in \S\ref{sec:foreground_bias}; we present lensing SNR forecasts for several CMB experiments in \S\ref{sec:forecasts} and summarize our results in \S\ref{sec:conclusion}.

Throughout this paper, we set the underlying cosmology to the results obtained from \planck{ } 2018 \mbox{TT,TE,EE+lowE+lensing} measurements \citep{planck18-6}.

%we assume a spatially flat $\Lambda$CDM Planck 2018 cosmology \citet{planck18-6} with Hubble constant $H_0 = 67.4 \text{ km s}^{-1}\text{Mpc}^{-1}$, baryon density $\Omega_b h^2 = 0.0224$, dark matter density $\Omega_c h^2 = 0.120$, matter density parameter $\Omega_m = 0.315$, primordial power spectrum with an amplitude $A_s = 2.101\times 10^{-9}$ and scalar spectral index $n_s = 0.965$, matter fluctuation amplitude $\sigma_8 = 0.811$, optical depth $\tau = 0.054$, and we assume a single massive neutrino with $m_\nu = 0.06 \text{ eV}$. 

%%%%%%%%%%%%%%%%%%%%%%%%%%%%%%%%%%%%%%%%
%%%%%%%%%%%%%%%%%%%%%%%%%%%%%%%%%%%%%%%%
%%%%%%%%%%%%%%%%%%%%%%%%%%%%%%%%%%%%%%%%

\section{CMB-Cluster Lensing}
\label{sec:theory}
CMB photons free-stream towards us from the surface of last scattering, constituting a diffuse source field that covers the entire sky. Due to the effect of gravitational lensing, the path of the CMB photons gets continuously deflected by the matter distribution between the last scattering surface and the observer, leading to a redistribution of the CMB temperature and polarization fields. Specifically, the fluctuation pattern of the lensed CMB field $X(\hat{\boldsymbol{n}}) \in [T(\hat{\boldsymbol{n}}), Q(\hat{\boldsymbol{n}}), U(\hat{\boldsymbol{n}})]$, with $T(\hat{\boldsymbol{n}})$ denoting the lensed temperature field, and $Q(\hat{\boldsymbol{n}})$ and $U(\hat{\boldsymbol{n}})$ denoting the lensed polarization fields, is given by a surface brightness conserving remapping of the unlensed field $\tilde{X}(\hat{\boldsymbol{n}})$ \citep{lewis06b}:
\begin{align}
    X(\hat{\boldsymbol{n}}) = \tilde{X}[\hat{\boldsymbol{n}} + \boldsymbol{\alpha}(\hat{\boldsymbol{n}})] \;, 
    \label{eq:cmb_lensing}
\end{align}
where $\hat{\boldsymbol{n}}$ denotes the line-of-sight direction and $\boldsymbol{\alpha}(\hat{\boldsymbol{n}})$ the deflection angle due to the mass distribution between the last scattering surface and the observer. 

In the case of lensing by a spherically symmetric halo, the deflection angle can be written as \refresponse{\citep{narayan96}}
 \begin{align}
 \boldsymbol{\alpha}(\hat{\boldsymbol{n}}) = - \frac{1}{\pi}\int\text{d}^2\hat{n}^\prime\kappa(\hat{\boldsymbol{n}}^\prime)\nabla\ln(|\hat{\boldsymbol{n}}-\hat{\boldsymbol{n}}^\prime|) \;,
 \label{eq:deflection_angle_isolated}
 \end{align}
where the convergence $\kappa$ is given by
\begin{align}
\kappa(\hat{\boldsymbol{n}}) = \frac{\Sigma(\hat{\boldsymbol{n}})}{\Sigma_\text{crit}} \;,
\label{eq:kappa}
\end{align}
with $\Sigma(\hat{\boldsymbol{n}})$ being the surface mass density of the cluster. The critical surface mass density $\Sigma_\text{crit}$ is a function of the angular diameter distance of the lens and source from the observer, and of the angular diameter distance between lens and source. 
Taking the divergence of Eq. (\ref{eq:deflection_angle_isolated}) leads to a simple relation between the deflection angle and the convergence:
 
\begin{align}
\refresponse{\nabla\cdot\boldsymbol{\alpha}(\hat{\boldsymbol{n}})} = - 2\kappa(\hat{\boldsymbol{n}}) \; .
\label{eq:kappa_alpha}
\end{align}

%The deflection angle is related to the convergence $\kappa(\boldsymbol{\theta})  =  \Sigma(\boldsymbol{\theta})/\Sigma_\text{crit}$ through

%\begin{align}
%nabla\boldsymbol{\alpha}  = 2\kappa \; .
%\label{eq:kappa_alpha}
%\end{align} 

%The critical mass density $\Sigma_\text{crit}$ is a function of the angular diameter distance of the lens and source from the observer, and of the angular diameter distance between lens and source. Since the convergence is given by the surface mass density of the lens, $\Sigma(\boldsymbol{\theta})$, divided by the critical lensing mass density $\Sigma_\text{crit}$, it represents a dimensionless surface mass density of the lensing system. 
When considering CMB lensing by galaxy clusters, only scales of a few arcminutes are of interest, corresponding to the angular size of galaxy clusters and to the amplitude of the produced deflection angle. On these small scales, the unlensed CMB is very smooth due to Silk damping \citep{silk68} and resembles a gradient field. Therefore, the lensed CMB field can be well approximated by a first order Taylor expansion of Eq. (\ref{eq:cmb_lensing}):

\begin{align}
 X(\hat{\boldsymbol{n}})  \approx  \tilde{X}(\hat{\boldsymbol{n}})+\refresponse{\boldsymbol{\alpha}(\hat{\boldsymbol{n}})\cdot\nabla \tilde{X}(\hat{\boldsymbol{n}})} \;.
\label{eq:cmb_lensing_approx}
\end{align}
Gravitational lensing will cause a local reversal of the background gradient, known as lensing dipole. As can be seen from Eq. (\ref{eq:cmb_lensing_approx}), the dipole signal depends linearly on the magnitude of the gradient for a given cluster mass. Hence, the highest SNR is obtained for clusters in front of a significant CMB background gradient. Since the root-mean-square (rms) temperature gradient is $ \sim 10\text{ }\mu\text{K}/\text{arcmin}$ and clusters can produce deflections of $\sim 1$ arcmin, the lensing signal in CMB temperature maps will be $\sim 10 \text{ }\mu\text{K}$. For polarization, the rms gradient is only $\sim 1\text{ }\mu\text{K}/\text{arcmin}$, leading to a signal of $\sim 1 \text{ }\mu\text{K}$ and therefore requiring $\times10$ lower noise levels compared to temperature data to be detected.

\section{Simulations}
\label{sec:simulations}
To analyze the performance of the lensing estimator, we generate simulated CMB temperature skies. %which, besides the CMB, include the lensing effect of the galaxy cluster, extragalactic foregrounds, the effects of the instrumental beam, as well as instrumental and atmospheric noise. 
Throughout this work, we neglect Galactic foregrounds since only small angular scales are relevant for CMB-cluster lensing. 

\subsection{Basic Ingredients}
The lensed CMB temperature power spectrum is computed with CAMB \citep{lewis00}.
%using the cosmological parameters from \citet{planck18-6} \maybe{Mention which parameter you are using exactly. Which column of the Table 2 from that paper}.
We create Gaussian realizations of the CMB power spectrum using the flat-sky approximation. The simulated maps have a size of $(60^{\prime} \times 60^{\prime})$ with a pixel resolution of $0.5^{\prime}$, which is large enough to avoid edge effects from discrete Fourier transformations, and to capture the large scale gradients across the map.  

%The concentration parameter is computed according to \citet{duffy08}:
%\begin{align}
% c(M, z) = A\left(\frac{M}{M_\text{pivot}}\right)^B(1+z)^C\; ,
% \label{eq:concentration}
%\end{align}
%with $M_\text{pivot} = 2\times10^{12}/h \text{ }M_\odot$, $h = 0.674$ \citep{planck18-6}, $A = 5.71$, $B = -0.084$, and $C = -0.47$. 

We model the cluster dark matter distribution using a Navarro-Frenk-White (NFW) profile \citep{navarro97}, which is characterized by its mass $M \equiv M_{200c} $ and redshift $z$. $M_{200c}$  is the mass inside a volume having a mean mass density equal to 200 times the critical density of the Universe at the redshift of the cluster. From the corresponding convergence profile \citep{bartelmann96} we compute the deflection angle using Eq. (\ref{eq:kappa_alpha}).
The cluster lensed map is obtained by remapping the CMB temperature anisotropy map using fifth order spline interpolation at the lensed positions. 

In real observations, the 
%\srini{\sout{cluster position might be miscentered due to normal positional uncertainties}}
cluster position depends on %the observable, namely 
the position of the brightest central galaxy, X-ray or SZ centroid, which, depending on the dynamical state of the cluster, can be different between observables. 
We study the bias due to the positional uncertainties in \S\ref{sec:positions}.
%To account for this 
To introduce the cluster centroid uncertainty in the simulated maps, we draw a positional offset from a Gaussian distribution centered around zero with a given standard deviation, $\sigma_\text{offset}$, and add it to the cluster position.

The maps are convolved by a Gaussian instrumental beam, which, in Fourier space, is given by

\begin{align}
     b_\ell =e^{-\frac{1}{2}\ell(\ell+1) \sigma^2}\;,
\end{align}
where $\sigma = \text{FWHM}/\sqrt{8 \ln(2)}$ and FWHM denotes the full width at half maximum of the beam. 

For the noise, we add a Gaussian realization of the following noise power spectrum model \citep{so19, abazajian19, choi20}: % \srini{SO is not the correct reference here. This representation dates back to first CMB papers in 1990s probably but SO introduced this weird Nred thing. Having SO is probably okay too.}

\begin{align}
     N_\ell = N_{\ell, \text{white}}+ N_\text{red}\left(\frac{\ell}{\ell_\text{knee}}\right)^{\alpha_\text{knee}}\;.
\end{align}
The first term on the right-hand sight refers to the white noise power spectrum of the detector of the experiment. The second term 
%\srini{\sout{describes the $1/f$ noise caused by fluctuations in the atmosphere}} 
is used to parameterize the atmospheric $1/f$ noise. It is the dominating noise source in ground based observations at large angular scales and will be added to the maps in \S \ref{sec:forecasts} to forecast the lensing SNR for different CMB experiments. As can be seen from Table \ref{tab_atm_noise_specs}, which contains the $\ell_\text{knee}$ and $\alpha_\text{knee}$ values for the considered experiments in this work, the atmospheric noise is not the main noise contributor, given that $\ell_\text{knee}$ is much smaller than the typical $\ell$-values for cluster-lensing.

\subsection{Extragalactic Foreground Signals}
\label{sec_clus_corr_foregrounds}
%\srini{\sout{To account for the kSZ and tSZ effects coming from the cluster acting as lens, we make use of \mdplsimname{} simulations \citep{omori2022}.}}

%The extragalactic foregrounds can be decomposed into emission from radio galaxies (RG), thermal emission from unresolved dusty star-forming galaxies at high redshifts, known as cosmic infrared background (CIB), and diffuse kSZ and tSZ signals. These are uncorrelated with the cluster and can therefore be included as Gaussian realizations using a foreground power spectra model based on SPT measurements \citep{george15, reichardt21}. 
%We do not lens any of the foregrounds. Although some part of the uncorrelated foregrounds will get lensed by the clusters, this effect is negligible and can therefore be neglected.  

%Additionally, we include the cluster-correlated kSZ and tSZ
%We include the cluster-correlated kSZ and tSZ signals to quantify the foreground-induced systematics in the lensing reconstruction. For this purpose, we make use of the \mdplsimname{} simulations \citep{omori2022}. We extract kSZ and Compton-y cutouts from \mdplsimname{} simulations
To quantify the impact of cluster-correlated kSZ and tSZ signals on the lensing reconstruction, we make use of \mdplsimname{} simulations \citep{omori2022}. 
To gain statistics, we extract kSZ and Compton-y cutouts %from \mdplsimname{} simulations 
for galaxy clusters in the mass range $M_{200c} \in [1.25, 1.69]\times10^{14}\text{ }M_\odot$ and redshift range $z\in[0.6, 0.8]$. 
The Compton-y cutouts are converted into tSZ cutouts in CMB temperature units using the relation
\begin{align}
    \Delta T_{\text{tSZ}} = \text{y}f(x)T_\text{CMB} \; ,
    \label{eq:comptony_to_tSZ}
\end{align}
with $T_\text{CMB} = 2.72548$ K \citep{fixsen09} being the mean CMB temperature;  $x = (h\nu)/(k_\text{B}T_\text{CMB})$ being the dimensionless frequency; and $h$ and $k_\text{B}$ being the Planck and Boltzmann constants, respectively. $f(x)$ describes the frequency dependence of the tSZ signal, which, ignoring relativistic corrections, is given by \citep{sunyaev81}
\begin{align}
     f(x) = x\coth(x/2)-4\;.
\end{align}
Ignoring the frequency dependent relativistic corrections is a reasonable assumption for the cluster masses considered in this work \cite{itoh98, chluba12}. %The kSZ and tSZ cutouts are added at the center of the lensed CMB temperature maps to analyze their impact on the lensing-based mass estimates (see \S \ref{sec:foreground_bias}). 
The kSZ and tSZ simulations are added at the center of the lensed CMB temperature maps to analyze their impact on the lensing-based mass estimates using single frequency tSZ maps at 150 GHz (see \S \ref{sec:foreground_bias}). Additionally, we use them to get realistic forecasts of the lensing SNR for different CMB experiments using optimally weighted tSZ maps based on 
all the available frequency bands of a given experiment (see \S \ref{sec_ilc} and \S \ref{sec:forecasts}).

For the lensing SNR forecasts, we also include the expected residual noise from cluster-uncorrelated extragalactic foregrounds (see \S \ref{sec_ilc}). The foregrounds include emission from radio galaxies (RGs), thermal emission from dusty star-forming galaxies making up the cosmic infrared background (CIB), and diffuse kSZ and tSZ signals. Since these foregrounds are uncorrelated with the cluster, they can be included as Gaussian realizations using a foreground power spectra model based on SPT measurements \citep{george15, reichardt21}\footnote{Although one can expect to observe galaxies within clusters, and as a result expect the signals from the CIB and RGs to be correlated with the cluster, we ignore them in this work for simplicity.}. 
\refresponse{We do not lens any of the foregrounds since this effect is negligible. To test the impact of lensing of the foregrounds, we create two mock datasets. In the first set, the foregrounds are lensed by the cluster, while in the second set, we add the foregrounds after lensing. We run the fitting procedure (see \S\ref{sec:results}) using the models that assume the foregrounds to be unlensed. The lensing masses inferred from the two sets differ by $\le 0.1\sigma$, and hence we ignore the effect of foreground lensing in the subsequent analysis.}

\subsection{Internal Linear Combination}
\label{sec_ilc}

%In \S\ref{sec:results} which involves pipeline validation (\S\ref{sec:validation});  comparing the performance of the estimator with MLE and QE (\S\ref{sec:comparison}); and biases from cluster mis-centering (\S\ref{sec:positions}) and foregrounds (\S\ref{sec:foreground_bias}), we simply use the 150 GHz map. 
%This map includes CMB; instrumental and atmospheric noise; and extragalactic foregrounds, both correlated and uncorrelated with the cluster, described above. The map is also convolved with a beam.

To forecast the lensing SNR for current and future experiments, we include information from all the frequency bands of a given experiment. We do this by using an internal linear combination (ILC) algorithm that preserves the signal of interest in an unbiased way while minimizing the variance of the output map (see e.g. \citep{tegmark96, tegmark03, cardoso08}). 
This is done by computing the optimal frequency dependent weights $\omega_{\ell}$ defined in Fourier space as %Harmonic space as 
\begin{align}
  \boldsymbol{\omega}_\ell = \frac{\boldsymbol{\mathrm C}_\ell^{-1} \boldsymbol{a}_\text{CMB}^{T}}{\boldsymbol{a}_\text{CMB}^{T} \boldsymbol{\mathrm C}_\ell^{-1}\boldsymbol{a}_\text{CMB}} \; ,
  \label{eq:ilc_weights}
\end{align}
where \refresponse{$\boldsymbol{a}_\text{CMB}^{T} = (1\text{ }...\text{ }1)_{N_\text{ch}\times1}$ is the mixing vector containing the spectral energy distribution of the CMB in temperature units in different frequency bands $\nu_i$}, $N_\text{ch}$ is the number of frequency channels of the experiment, and $\refresponse{\boldsymbol{\mathrm C}_\ell}$ is the $N_\text{ch} \times N_\text{ch}$ covariance matrix containing the auto- and cross-spectra \refresponse{$C_{\ell}^{(\nu_i, \nu_j)}$} of maps from different channels at a given multipole $\ell$. 
The covariance matrix receives contribution from \refresponse{the sky signals and the experimental noise in the map, such that}
\begin{align}
C_\ell^{(\nu_i, \nu_j)} = C_{\ell, \text{RGs}}^{(\nu_i, \nu_j)} + C_{\ell, \text{CIB}}^{(\nu_i, \nu_j)} + C_{\ell, \text{kSZ}}^{(\nu_i, \nu_j)} + C_{\ell, \text{tSZ}}^{(\nu_i, \nu_j)} + C_{\ell, \text{tSZ-CIB}}^{(\nu_i, \nu_j)} + N_{\ell}^{(\nu_i, \nu_j)}
 \; .
  \label{eq:ilc_cov}
\end{align}
\refresponse{Besides these individual terms, we also include the additional variance that arises due to the cross-correlation of the tSZ and CIB signals, $C_{\ell, \text{tSZ-CIB}}$ \citep{addison12, reichardt21}.}
We model the power spectra of \refresponse{the} extragalactic components based on the best-fit power spectra at 150 GHz from \citet{george15, reichardt21}.
The residual foreground and noise spectra in the ILC map can be computed as
\begin{align}
	\boldsymbol{N}_{\ell, \text{ILC}} = \frac{1}{\boldsymbol{a}_\text{CMB}^{T} \boldsymbol{\mathrm C}_\ell^{-1}\boldsymbol{a}_\text{CMB}} \; .
    \label{eq:residual}
\end{align}

\section{Lensing Estimator}
\label{sec:estimator}

\begin{figure}[!htb]
 \centering
 \includegraphics[width=\textwidth,keepaspectratio]{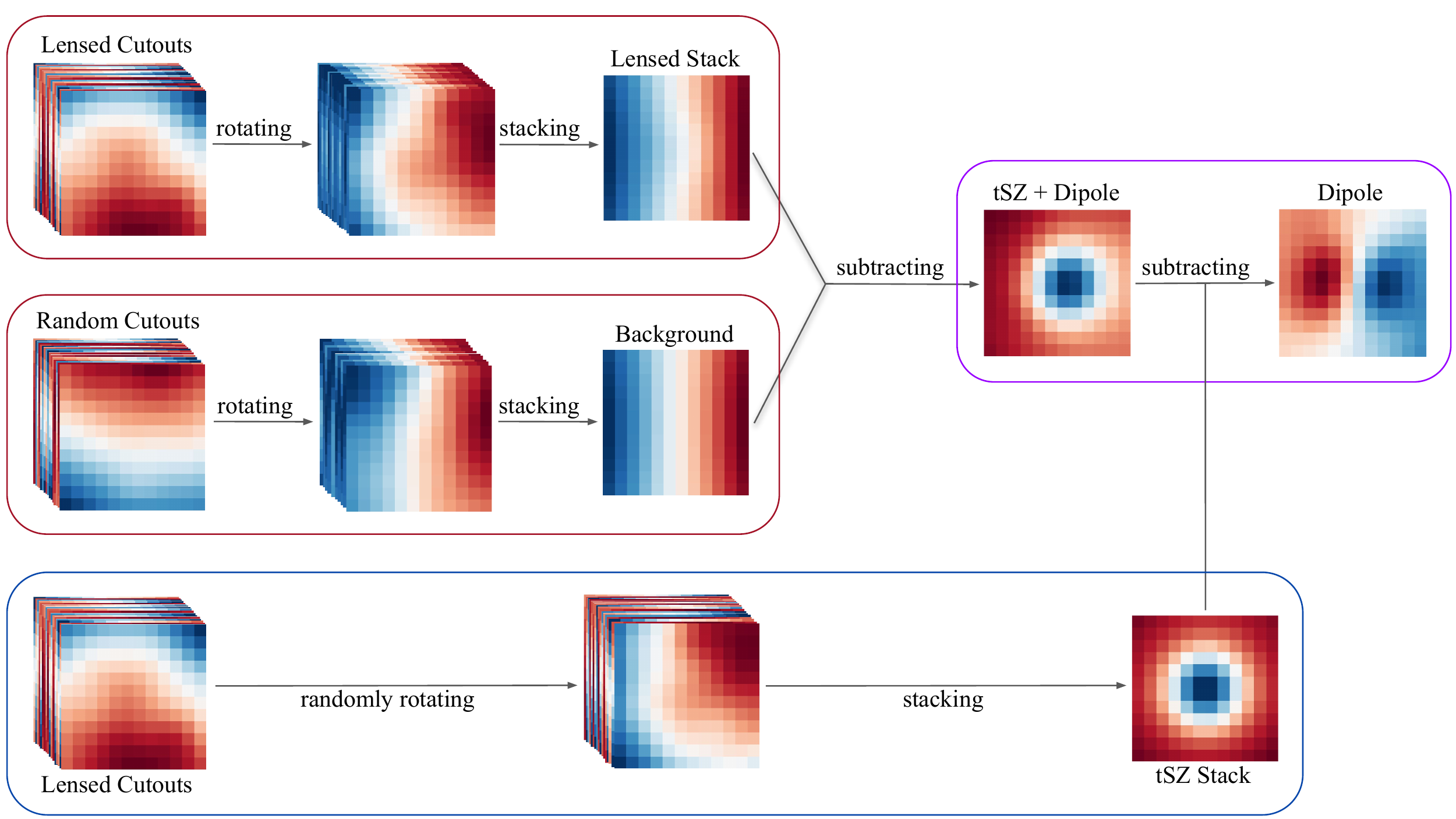}
 \caption{Illustration of the lensing pipeline when applied to CMB temperature mock data. Besides the lensing signal, the cluster-centered cutouts include the cluster-correlated kSZ and tSZ signals. All the cutouts have been smoothed by a Gaussian beam with FWHM = $1'$ and contain a white noise level $\Delta T_\text{white} = 2$ $\mu$K-arcmin. While the kSZ effect cancels out during the stacking process, the rotational-invariant tSZ signal remains in the cluster-lensed stack. An estimate of the mean tSZ signal can be obtained by randomly rotating and stacking the lensed cutouts. This will remove the CMB gradients and lensing dipoles, and thus only leaving the mean tSZ signal. The random rotation process ensures that the gradients between different cutouts are uncorrelated. Additionally, this process can be repeated to obtain a given number of tSZ stacks which can be averaged over to reduce the residual noise within the tSZ stack. Note that, when using CMB polarization data, the additional tSZ mitigation step is not necessary, since foregrounds are largely unpolarized.
 }
\label{fig:pipeline_tsz_mitigation}
\end{figure}

Ignoring the effect of foregrounds, an estimate of the mean lensing dipole signal can be obtained from $N_\text{clus}$ cluster-centered and $N_\text{rand}$ random CMB maps using the following steps \citep{raghunathan19c}:
%An estimate of the mean dipole signal can be obtained from a sample of cluster-lensed CMB cutouts using the following steps \citep{raghunathan19c}:

\begin{enumerate}
	\item Using the central $(6^{\prime} \times 6^{\prime})$ region in every map, %simulated map with size $(60^{\prime} \times 60^{\prime})$
    compute the median gradient direction, \refresponse{\mbox{$\theta_{\nabla T} = \tan^{-1}(\nabla_yT/\nabla_xT)$}}, and magnitude,  \refresponse{\mbox{$|\nabla T| = \sqrt{\nabla_xT^2+\nabla_yT^2}$}}.
	\item Rotate each map along its gradient direction.
	\item Extract central $(6^{\prime} \times 6^{\prime})$ cutouts from the rotated maps.
	\item Subtract the median from each cutout.
	\item Compute the gradient magnitude weighted cluster-lensed and background stacks, $\boldsymbol{\mathrm s}_\text{clus}$ and $\boldsymbol{\mathrm s}_\text{bg}$, respectively. 
	\item Get an estimate of the mean lensing dipole $\boldsymbol{\mathrm s}_\text{dipole}$ by subtracting the background stack from the cluster-lensed stack.
\end{enumerate}
%Note that, in contrast to \citetalias{raghunathan19c}, we first rotate the maps and then extract the cutouts instead of rotating the cutouts themselves. This is done to avoid artifacts at the edges of the cutouts due to the rotation process. Additionally, the cutout size used in this work has a size of $(6^{\prime} \times 6^{\prime})$, instead of $(10^{\prime} \times 10^{\prime})$, which was used in \citetalias{raghunathan19c}. Using this smaller cutout size does not significantly affect the SNR since nearly everything beyond the $(6^{\prime} \times 6^{\prime})$ region is dominated by noise. \\
Note that the cutout size used in this work is slightly smaller compared to the size used in the original paper: $(6^{\prime} \times 6^{\prime})$ vs $(10^{\prime} \times 10^{\prime})$. 
%\sout{Using this smaller cutout size does not significantly affect the SNR since nearly everything beyond the $(6^{\prime} \times 6^{\prime})$ region is dominated by noise.}
Using this smaller cutout size does not degrade the SNR since the majority of the lensing signal comes from regions close to the center. 
Additionally, the smaller cutout size reduces the number of elements in the covariance matrix.
%Additionally, we first rotate the maps and then extract the cutouts instead of rotating the cutouts themselves to avoid artifacts at the edges of the cutouts due to the rotation process. 
In summary, the individual stacks are given by 

\begin{align}
\boldsymbol{\mathrm s}_\text{clus} & = \frac{\sum_{c}^{N_\text{clus}}w_c\left[\tilde{\boldsymbol{\mathrm d}}_{c} - \langle\tilde{\boldsymbol{\mathrm d}}_{c}\rangle\right]}{\sum_{c}^{N_\text{clus}}w_c}\; ,
\label{eq:stack_clus}
\\
\boldsymbol{\mathrm s}_\text{bg} & = \frac{\sum_{r}^{N_\text{rand}}w_r\left[\tilde{\boldsymbol{\mathrm d}}_{r} - \langle\tilde{\boldsymbol{\mathrm d}}_{r}\rangle \right]}{\sum_{r}^{N_\text{rand}}w_r}  \; ,
\label{eq:stack_bg} 
\\
\boldsymbol{\mathrm s}_\text{dipole} & = \boldsymbol{\mathrm s}_\text{clus}  - \boldsymbol{\mathrm s}_\text{bg} \; .
\label{eq:stack_dipole}
\end{align}
Every rotated cutout, $\tilde{\boldsymbol{\mathrm d}}_{i}$, is weighted by the corresponding gradient magnitude, $\refresponse{w_i = |\nabla T|_i}$, since the dipole signal scales linearly with the CMB background gradient for a given cluster mass. When using real observations, additional weights can be used based on the inverse noise variance $\sigma_i^{-2}$ in each cutout, giving a final weight of $\refresponse{w_i = |\nabla T|_i\sigma_i^{-2}}$. Since the noise realizations in our CMB simulations are Gaussian, the noise weights will be the same for each cutout and are neglected in this work. 

When working with CMB temperature data, cluster-correlated foregrounds, such as the clusters' own kSZ and tSZ signals, can significantly bias the mass results. Since the estimator analyzed in this work is based on stacking rotated cutouts, the kSZ effect should not compose a source of bias, as the radial peculiar velocity of a cluster can be positive or negative. The tSZ signal, on the other hand, is rotationally invariant on average, and will be the dominant signal in the background subtracted stack if not accounted for. An estimate of the mean tSZ signal can be obtained from randomly rotated cluster-lensed cutouts, $\tilde{\boldsymbol{\mathrm d}}_{c}^{\theta_\text{rand}}$, by computing the corresponding gradient weighted stack. The random rotation ensures that any correlations between CMB gradients in different cutouts, which could exist between cluster-centered cutouts that are close to each other on the sphere, will be eliminated. Therefore, the resulting stack will only contain the mean tSZ signal and some residual noise, which can be reduced by averaging over a given number of tSZ stacks obtained from different sets of random rotations:
\begin{align}
  \boldsymbol{\mathrm s}_\text{tSZ} &= \left \langle \frac{\sum_{c}^{N_\text{clus}}w_c\left[\tilde{{\boldsymbol{\mathrm d}}}_c^{\theta_{\text{rand}}}-\langle{\tilde{\boldsymbol{\mathrm d}}}_c^{\theta_{\text{rand}}}\rangle\right]}{\sum_{c}^{N_\text{clus}}w_c} \right \rangle_{\theta_{\text{rand}}} \; .
  \label{eq:stack_tsz}
\end{align}
\addcomments{While we do not explicitly show, we note that the above technique of removing the tSZ signal will also help in removing other cluster-correlated foreground signals, like the emission from member galaxies.}
The final dipole stack is then given by:
\begin{align}
 \boldsymbol{\mathrm s}_\text{dipole} = \boldsymbol{\mathrm s}_\text{clus} - \boldsymbol{\mathrm s}_\text{bg} - \boldsymbol{\mathrm s}_\text{tSZ}\; .
  \label{eq:stack_dipole_tsz}
\end{align}
Fig. \ref{fig:pipeline_tsz_mitigation} illustrates the above steps to extract the lensing dipole from CMB temperature maps.

To reduce the noise penalty in the estimation of the median value of the gradient direction and magnitude, a Wiener filter of the form 
\begin{align}
W_\ell = 
\begin{cases}
\frac{C_\ell}{C_\ell + N_{\ell}} &  ,\text{ } \ell \leqslant 2000 \\
0 &  ,\text{ } \ell > 2000
\end{cases} \;  
\label{eq:filter}
\end{align}
is applied to the lensed and random maps. $C_\ell$ is the lensed CMB temperature power spectrum and $C_{\ell, \text{noise}}$ refers to the total noise power spectrum of the map. The sharp multipole cut at $\ell = 2000$ is used to remove the lensing signal in the cluster-lensed cutouts, which magnifies the background image and leads to a decrease of the CMB gradient \citep{hu07}.
%This should ensure that the random and cluster-lensed cutouts are rotated and weighted in the same way, since most of the CMB gradient comes from multipoles $\ell \leqslant 2000$ due to Silk damping of the acoustic peaks. 
%For clusters used in this work, that span roughly few arcminutes, this cut is sufficient. For massive low redshift clusters, this cut can be shifted to lower multipoles to remove the effect of lensing \citep{hu07}.
Note that this cut does not degrade the SNR on the gradient measurement since the majority of the CMB gradient comes from $\ell \leqslant 2000$ and the modes beyond those scales are exponentially suppressed due to Silk damping.
Thus, this $\ell$-cut should ensure that the random and cluster-lensed cutouts are rotated and weighted in the same way.
We compute the gradient within the $(6^{\prime} \times 6^{\prime})$ box around the center of the filtered maps using second order accurate central differences. 
% Fig. \ref{fig:pipeline_baseline} summarizes the above steps to obtain the lensing dipole signal when foregrounds are negligible. \\

%\begin{figure}
%\includegraphics[width=1\textwidth, keepaspectratio]{figures/pipeline.png}
% \centering
% \includegraphics[width=15.5cm , height=6.cm]
% {figures/lensing_pipeline_baseline.eps}
 %{figures/cluster_lensing_pipeline.pdf}
% \caption{Lensing pipeline applied on foreground-free CMB simulations, smoothed by a beam with FWHM = 1$^\prime$ and containing a white-noise level of 2 $\mu$K-arcmin. The stacks have been computed from 2,500 lensed a 50,000 unlensed and sims. The cluster mass and redshift used for the lensed cutouts are $M = 3\times10^{14} \text{ }M_\odot$ and $z = 0.7$, respectively. The dipole is obtained by subtracting the background stack from the cluster lensed one. %This pipeline can be used when astrophysical foregrounds can be neglected, which is for example the case when using CMB polarization data since foregrounds are largely unpolarized.
 %}
%\label{fig:pipeline_baseline}
%\end{figure}

%%%%%%%%%%%%%%%%%%%%%%%%%%%%%%%%%%%%%%%%%%%%%%%%
%%%%%%%%%%%%%%%%%%%%%%%%%%%%%%%%%%%%%%%%%%%%%%%%
%%%%%%%%%%%%%%%%%%%%%%%%%%%%%%%%%%%%%%%%%%%%%%%%
\section{Statistical and Systematic Analysis}
\label{sec:results}
%We now perform several statistical and systematic error checks and compare the performance of the estimator to standard CMB-cluster lensing estimators. For the following sections, we use a total of $25,000$  lensed cutouts, split into $N_\text{sets} = 10$ sets, each containing $N_\text{clus} = 2,500$ cutouts. Splitting the cluster-lensed cutouts into individual sets decreases the computational time for the model generation and enables us to highlight the distribution of the individual likelihood curves around the input mass (see \S\ref{sec:validation}). We use a redshift of $z = 0.7$ for all the clusters. The cluster masses, white noise levels, as well as the inclusion of positional offsets and foregrounds are described in the individual sections. We use $N_\text{rand} = 50,000$ random cutouts for the background stack. The large number of random simulations is used to ensure that the variance in the background stack is negligible. All the simulation have been smoothed by a Gaussian instrumental beam with FWHM = $1^{\prime}$.  
We now perform several statistical and systematic error checks and compare the performance of the estimator to standard CMB-cluster lensing estimators. For these checks, we use $N_\text{clus}=25,000$ cutouts for the lensed stack and $N_\text{rand} = 50,000$ random cutouts for the background stack. We set the redshift to $z = 0.7$ for all the clusters. The large number of random maps is used to ensure that the variance in the background stack is negligible. 
The assumed cluster masses, white noise levels, positional offsets and foregrounds are described below in the individual sections. All the simulations have been smoothed by a Gaussian instrumental beam with FWHM = $1^{\prime}$.  

The cluster mass is estimated from the lensing dipole stack by computing the natural logarithm of the likelihood function:
\begin{align}
{\mathrm \ln} \mathcal{L}(M|\boldsymbol{\mathrm s})  =  -\frac{1}{2} \left[ \boldsymbol{\mathrm s}-\boldsymbol{\mathrm m}(M) \right]^T \boldsymbol{\mathrm C}^{-1} \left[ \boldsymbol{\mathrm s}-\boldsymbol{\mathrm m}(M) \right]  \; ,
 \label{eq:likelihood}
\end{align}
where $\boldsymbol{\mathrm s}$ is the mock dipole vector and $\boldsymbol{\mathrm m}(M)$ the mass-dependent model vector. The pixel-pixel covariance matrix $\boldsymbol{\mathrm C}$ is estimated using a jackknife re-sampling technique by dividing the $N_\text{clus}$ mock observations into $N_{\text{jk}}$ sub-samples: 
\begin{align}
\boldsymbol{\mathrm C} = \frac{N_{\text{jk}}-1}{N_{\text{jk}}}\sum_{i = 1}^{N_{\text{jk}}} (\boldsymbol{\mathrm s}_i- \langle \boldsymbol{\mathrm s} \rangle)(\boldsymbol{\mathrm s}_i- \langle \boldsymbol{\mathrm s} \rangle)^{T}\;,
 \label{eq:covariance}
\end{align}
\refresponse{where $\boldsymbol{\mathrm s}_i$ is the lensing dipole vector of the $i^{th}$ sub-sample and $\langle \boldsymbol{\mathrm s} \rangle$ the average dipole vector of all the sub-samples.}

The dipole models are generated using a flat prior for masses selected from a parameter grid  ranging from $M \in [0, 6] \times 10^{14}\text{ }M_\odot$ with a mass resolution of $\Delta M = 0.01 \times10^{14}\text{ }M_\odot$. The model for each mass is obtained from $N_\text{clus}$ fixed CMB simulations, lensed by the cluster with the respective mass. We smooth the maps with a Gaussian beam identical to the one used for the mock data. The background stack for the models is obtained from the corresponding unlensed simulations. We subtract this stack from each lensed stack to get the final lensing dipole models. %As was done in \citetalias{raghunathan19c}, 
We add the same noise level to the model simulations that was added to the mock data when estimating the median gradient direction and magnitude. Otherwise, the uncertainty in the determination of the gradient directions will be smaller for the models than for the mock data, leading to a higher amplitude in the model dipoles since the cutouts will be aligned more precisely, and thus to a bias towards lower mass values. 

Note that, to create the models, we do not rotate each of the lensed maps by the gradient angles obtained from the corresponding Wiener filtered cutouts, since this introduces a scatter in the likelihood curves. This scatter is due to differences in the uncertainties in the gradient estimation for different cluster masses. Instead, we lens the $i^{th}$ CMB simulation by every mass in the mass bin, infer the gradient angle for each of these lensed simulations, and rotate the maps by the median angle. This ensures that all the different lensed model stacks have been rotated in the same way. 

%All the mass estimates quoted in this paper are are obtained from the combined likelihood of all the individual sets:
%\begin{align}
%{\mathrm \ln} \mathcal{L}_\text{comb}(M|\boldsymbol{\mathrm s})  = \sum_{i=0}^{N_\text{sets}}  \mathcal{L}(M|\boldsymbol{\mathrm s}_i) \; .
%\label{eq:likelihood_comb}
%\end{align}
%Specifically, the best-fit mass, $M_\text{fit}$, and corresponding $1\sigma$ uncertainty, $\Delta M$, are given as 50th percentile and half the difference between the 16th and 84th percentile of the combined likelihood function, respectively. 
The best-fit mass, $M_\text{fit}$, and corresponding $1\sigma$ uncertainty, $\Delta M$, are given as 50th percentile and half the difference between the 16th and 84th percentile of the likelihood function, respectively.

%If not stated otherwise, the maps contain a white noise level of $\Delta T_\text{white} = 2\text{ }\mu$K-arcmin. 
%None of the simulations does include cluster-uncorrelated extragalactic foregrounds as these are mitigated using multifrequency observations and thus depend on the specific experiment. We consider this case in \S\ref{sec:forecasts} for different CMB experiments. \\

\subsection{Pipeline Validation}
\label{sec:validation}
\begin{figure}
    \centering
    \includegraphics[width=\textwidth,keepaspectratio]{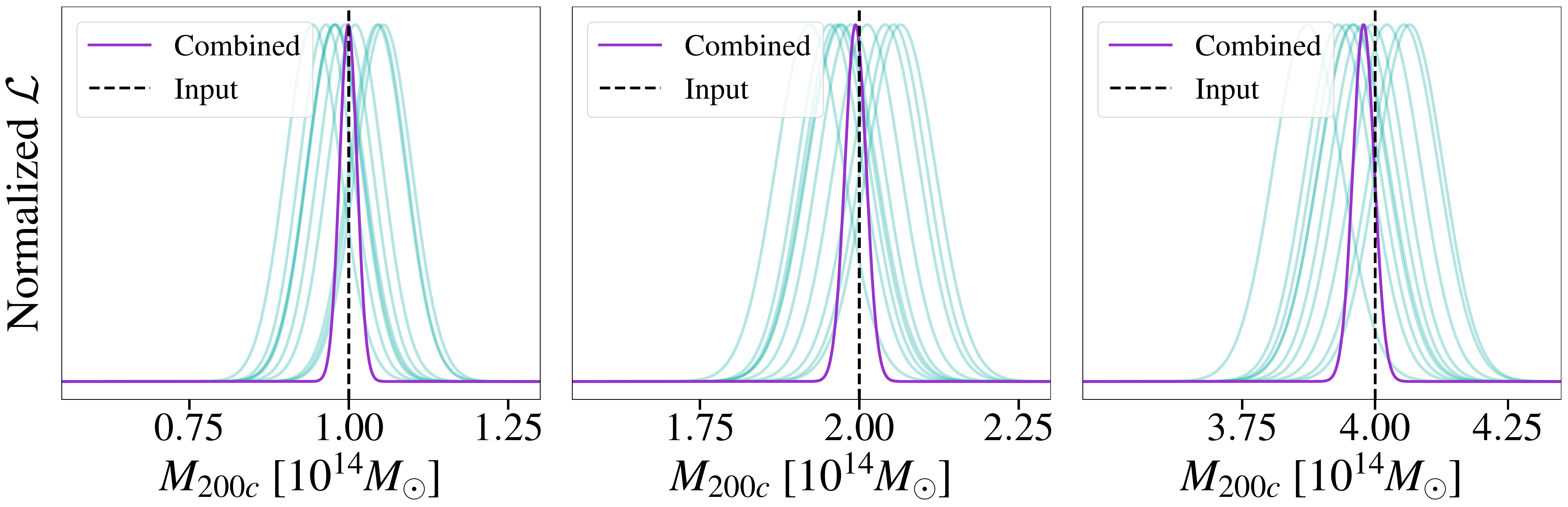}
    \caption{Individual likelihood curves (light shaded turquoise curves) from 10 sets for three mass cases: $M_\text{input} \in [1, 2, 4]\times10^{14}\ \msol$. Each set includes 2,500 lensed maps, smoothed by a Gaussian beam with FWHM = $1^{\prime}$ and containing a map noise level of $\Delta T_\text{white} = 2$ $\mu$K-arcmin. The combined likelihood curve of the 10 sets is shown as solid purple curve. The input mass used for the mock data is highlighted as dashed black line.}
\label{fig:pipeline_validation}
\end{figure}

We begin our analysis by applying the estimator to simulations lensed by different masses to verify whether we can properly recover the input masses. Besides the lensing signal and the instrumental beam, the simulations contain a white noise level $\Delta T_\text{white}$ = 2 $\mu$K-arcmin, which roughly corresponds to the noise expected for \spt{} \citep{bender18} and \cmbsfourwide{} \citep{abazajian19} experiments. We consider three input masses: $M_\text{input} \in [1, 2, 4]\times10^{14}\text{ }M_\odot$, for which we compute the lensing dipole stacks according to Eq. (\ref{eq:stack_dipole}). 
The likelihood curves of 10 individual simulation sets, each containing 2,500 clusters, can be seen as light shaded curves in Fig.~\ref{fig:pipeline_validation} for each mass case. %, together with the combined likelihood of the 25,000 clusters. 
The combined likelihood of the 25,000 clusters is shown as thick purple curve. 
We find median masses and 1$\sigma$ uncertainties of $M_{1\times10^{14}M_\odot} = (1.00 \pm 0.01)\times10^{14} \text{ }M_\odot$, $M_{2\times10^{14}M_\odot} = (1.99 \pm 0.02)\times10^{14}\text{ }M_\odot$, and $M_{4\times10^{14}M_\odot} = (3.98\pm 0.02)\times10^{14}\text{ }M_\odot$, indicating that, although we can recover the input masses within $1\sigma$ for the considered settings, there seems to be a small systematic shift towards lower masses with increasing cluster mass. The reason for this being that the filter given by Eq. (\ref{eq:filter}) does not entirely remove the lensing effect of the cluster, making the rotation process slightly mass dependent. We verify this by redoing the analysis, using the  gradient angles estimated from the corresponding underlying unlensed simulations to rotate the lensed mock data and model simulations, and see no mass-depend shift in the recovered masses.
%\srini{Similar to cluster lensing, the cluster-correlated foreground signals can also cause issues in the gradient measurement.}

\addcomments{Besides using a smaller value for the multipole cut for the Wiener filter given by Eq. (\ref{eq:filter})}, a possible method to get unbiased masses for massive clusters would be to inpaint the maps \citep{benoit13, raghunathan19b} before applying the Wiener filter. Inpainting mitigates the lensing signal, as well as any cluster-correlated foreground signals, by masking the cluster region and filling the corresponding pixel values based on information from surrounding regions using constrained Gaussian realizations. %In that case, it could be possible to use the gradient angles of the corresponding unlensed simulations to create the model lensing dipole stacks without introducing a low-mass bias. 
While outside the scope of this work, inpainting the maps before estimating the CMB background gradient orientation constitutes an interesting implementation for a future analysis. 

%the filter given by Eq. (\ref{eq:filter}) does not entirely remove the lensing effect of the cluster, making the rotation process slightly mass dependent. This leads to a mass depend systematic shift to lower masses which increases with the cluster mass since the uncertainties of the gradient orientation estimation increase with the lensing signal. Since we use the gradient orientations and weights from the unlensed model simulations, the models are aligned more precisely and therefore have a higher dipole amplitude, thus similar to the case when we do not include the same noise level to the model simulations for the gradient estimation. Using the gradient orientations and weights from the lensed model simulation would reduce the dipole amplitudes in a similar way as for the mock data, which is how the models were created in \citetalias{raghunathan19c}. However, using this approach for the model creation, we see a scatter in the likelihood curves, since the uncertainties in the gradient angle determination will be different for each model, making the results unreliable. A possible way to reduce the bias is to inpaint the maps before applying the Wiener filter, since inpainting should completely remove the lensing signal as well as any foreground signals, leaving only the CMB gradient. %Since the typical mean mass of a cluster sample is $\sim 2\times10^{14}M_\odot$ and map noise levels for current and upcoming CMB surveys exceed 2 $\mu$K-arcmin, inpainting the maps will not be necessary.

\subsection{Estimator Comparison}
\label{sec:comparison}
\begin{figure}
    \centering
    \includegraphics[width=12cm , height=8cm]{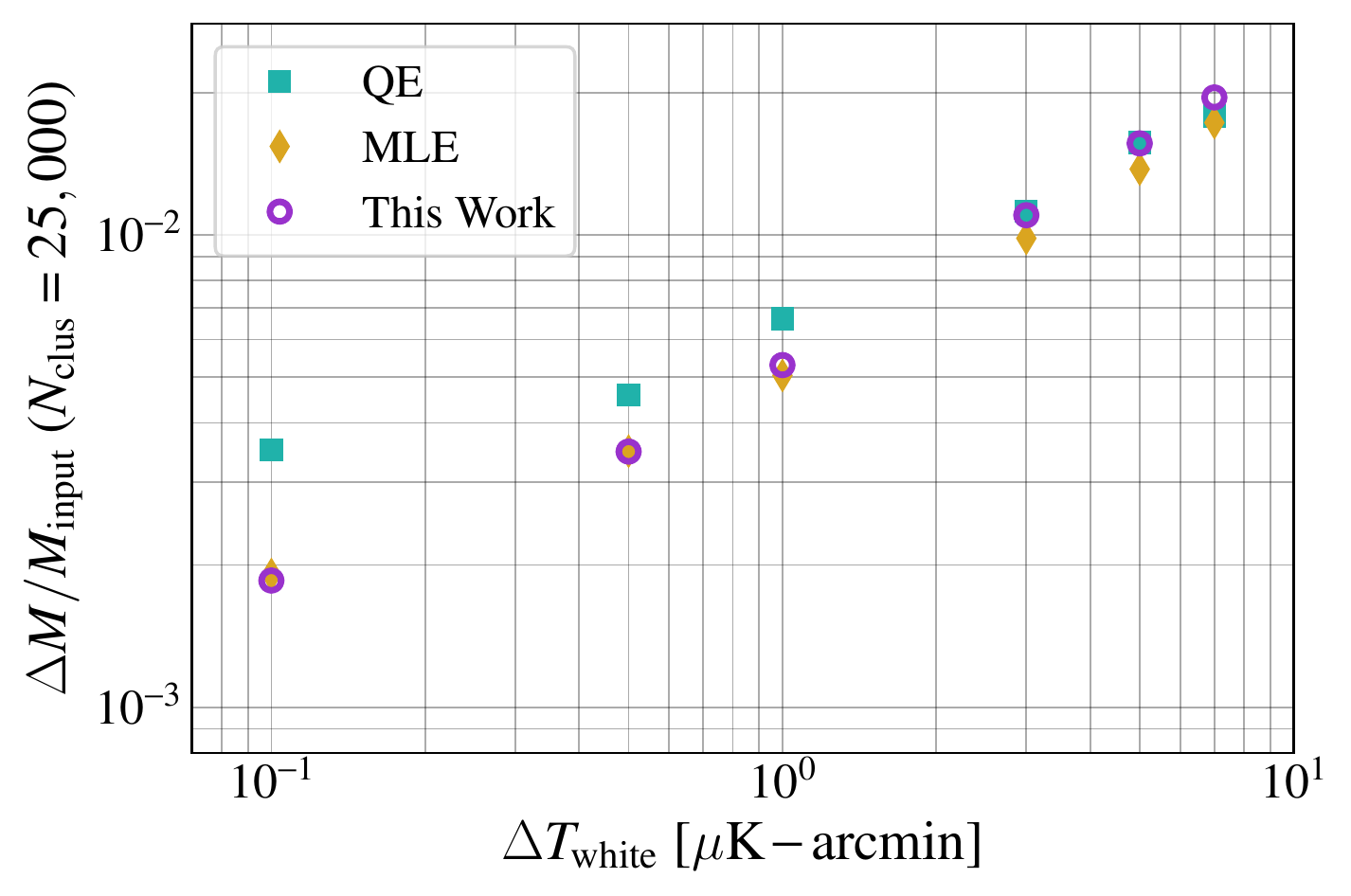}
    \caption{Fractional mass uncertainties for white noise levels $\Delta T_\text{white} \in [0.1, 0.5, 1, 3, 5, 7]\text{ }\mu$K-arcmin using the QE (turquoise squares), the MLE (yellow diamonds), and the estimator of this work (purple circles). 
    The cluster stack is obtained from $N_{\rm clus} = 25,000$ lensed cutouts.
    The three considered estimators have a similar performance for $\Delta T_\text{white} \geqslant 3.0$ $\mu$K-arcmin.
    For lower noise levels, both, the MLE and our estimator, outperform the QE. Specifically, for the case $\Delta T_\text{white} = 0.1$ $\mu$K-arcmin, the fractional mass uncertainty of the QE is $\times 2$ worse.
    }
\label{fig:estimator_comparison}
\end{figure}

In this section, we compare the fractional mass uncertainty, $\Delta M/M_\text{input}$, of the lensing estimator to those obtained for the MLE \citep{seljak00b, dodelson04, lewis06a}  and QE \citep{hu07, raghunathan17b}.
The QE estimates the mean cluster mass by exploiting the lensing-induced correlations between previously uncorrelated
modes. %By taking a radial profile of the stacked convergence map, it is possible to recover the mean cluster mass of the sample by fitting an NFW convergence profile to the data profile. 
In the MLE, cluster-lensed CMB templates are fitted to observed CMB maps using the full pixel-space likelihood. For easier comparison, we use the same settings as in \citet{raghunathan17b}: \mbox{$M = 2\times10^{14}\text{ }M_\odot$}, $z = 0.7$, FWHM = 1$^{\prime}$, and $\Delta T_\text{white} \in [0.1, 0.5, 1, 3, 5, 7]\text{ }\mu$K-arcmin. Since the values in \citet{raghunathan17b} were obtained for $N_\text{clus} = 100,000$, we scale those values by a factor of $\sqrt{100,000/25,000} = 2$. The resulting fractional mass uncertainties can be seen in Fig. \ref{fig:estimator_comparison}, together with the values obtained in \citet{raghunathan17b} for the QE and MLE. 

The MLE and the estimator analyzed in this work give similar fractional mass uncertainties over the whole noise range. The QE has a similar performance for white noise levels $\Delta T_\text{white} \geqslant 3.0$ $\mu$K-arcmin but is outperformed for lower noise levels. For the lowest considered noise level, $\Delta T_\text{white} = 0.1$ $\mu$K-arcmin, the fractional mass uncertainty of the MLE and the current estimator are improved by $\times2$ compared to the one obtained for the QE. %, similar to the results obtained in \citet{raghunathan17b}. 
Since the QE is only a linear approximation, % to the MLE \citep{hirata03}, 
it misses some of the information included in the MLE and the estimator of this work. 
%Using an iterative version for the QE \citep{yoo08, yoo10} makes it possible to recover the extra information that is missing in the standard QE of \citet{hu07}, making the QE compatible with the MLE and the current estimator.
As demonstrated by \citep{yoo08, yoo10}, using an iterative version of the QE can make it match with the MLE and the current estimator even for low map noise levels.
%Additionally, the simulations used for this analysis do not contain any foregrounds, which will set an effective noise floor at the $10^{-2}$ level %\srini{hmm, you mean mass uncertainty. Better to say that the constraints saturate after a particular noise level. Probably just refer to right panel Figure from the MLE paper.}.
%The fractional mass uncertainties for our estimator are comparable to the one of the MLE for high white noise levels. For low noise levels ($\Delta T_\text{white} \leq 1 \text{}\mu$K), the uncertainties converge around a value of $\Delta M/M \approx 3\times10^{-3}$. This limit is due to the uncertainties in the gradient estimation which for these low noise levels are dominated by the CMB fluctuation rather than the white noise. However, real CMB observations will contain additional noise due to the atmosphere (for ground based experiments) and astrophysical foregrounds. While astrophysical foreground contamination can be reduced, there will always be a residual left which will set an effective noise floor for CMB temperature-based lensing reconstruction. This noise floor will limit the fractional mass uncertainties in temperature to $\sim 10^{-2}$ (see right panel of Fig. 2 in \citet{raghunathan17b}) and thus make our estimator compatible with the QE and MLE for real CMB temperature observations. 

\subsection{Systematic Biases and Mitigation Strategies}
\label{sec:sys_bias}
We now turn to the systematic analysis by examining biases due to uncertainties in the cluster positions and the clusters' own tSZ and kSZ signals. The relative bias due to these systematics is quantified as

\begin{align}
    \bias = \frac{M_\text{fit}}{M_\text{input}}-1 \; .
\end{align}

\subsubsection{Cluster Positions}
\label{sec:positions}

\begin{figure}
    \centering
    \includegraphics[width=12cm , height=8cm]{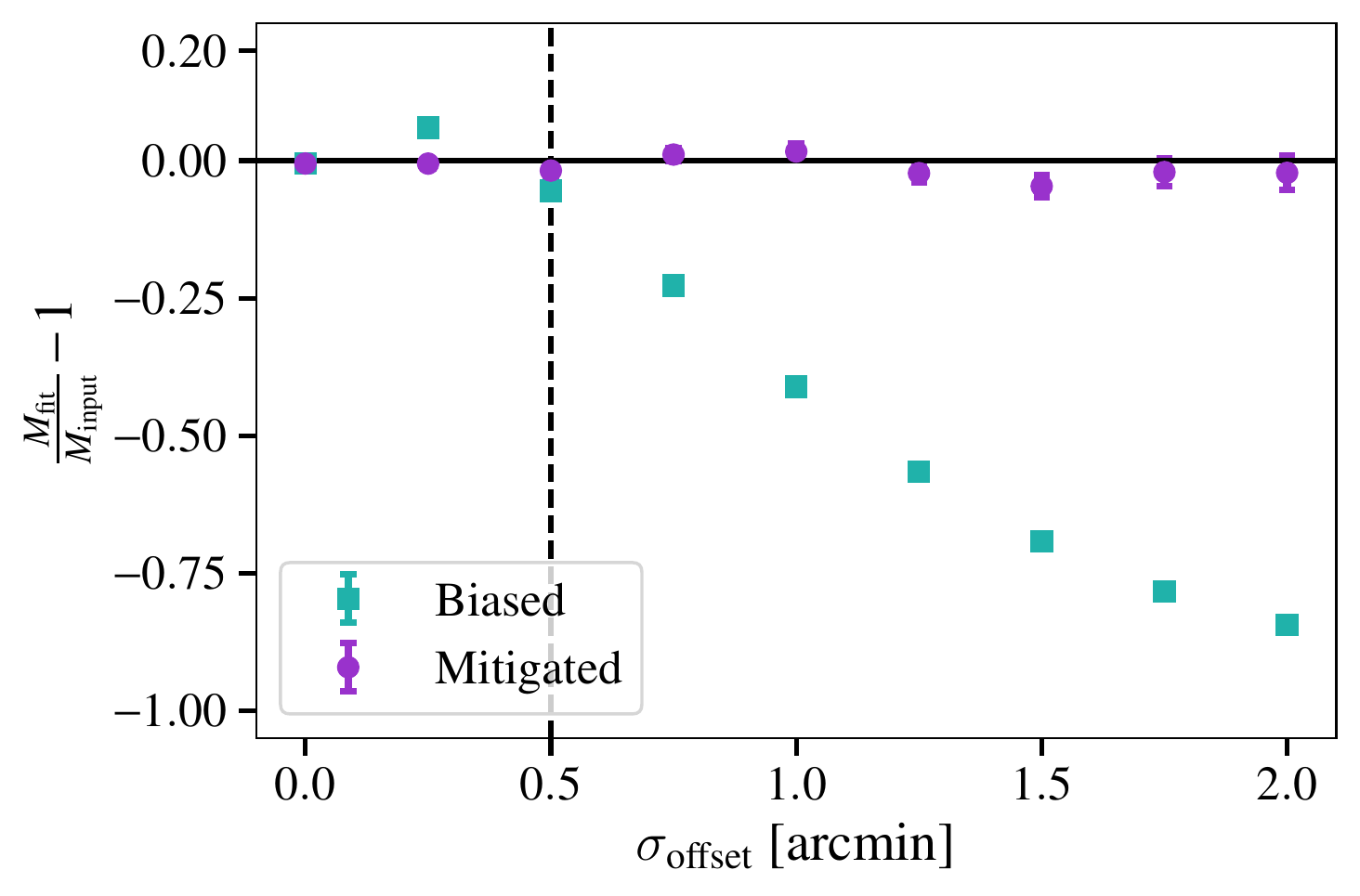}
    \caption{Bias analysis due to uncertainties in the cluster positions for 25,000 clusters. Unaccounted positional uncertainties in the data will lead to a bias towards lower masses (turquoise squares). For a typical positional offset of $0.5^{\prime}$ expected in SZ surveys \citep{song12}, marked with a dashed black line, we find a bias $\bias =  -0.05\pm0.01$. 
    %Including the same level of uncertainties 
    Accounting for the uncertainties in the models helps mitigating the bias at the cost of an increased mass error (purple points), which is $\lesssim 10\%$ for $\sigma_\text{offset} \lesssim 0.5^{\prime}$, and hence negligible.
    }
\label{fig:cluster_positions}
\end{figure}

To quantify the potential bias due to normal positional uncertainties in the cluster positions, we add a random offset to the position of each cluster, taken from a Gaussian distribution, $N(0, \sigma_\text{offset})$, with a standard deviation $\sigma_\text{offset} \in [0', 0.25', 0.5^{\prime}, 0.75^{\prime}, 1^{\prime}, 1.25^{\prime}, 1.5^{\prime}, 1.75^{\prime}, 2^{\prime}]$. We use a cluster mass $M = 2\times10^{14}\text{ }M_\odot$ and a white noise level $\Delta T_\text{white} = 2$ $\mu$K-arcmin. Fig.~\ref{fig:cluster_positions} shows the biases due to the centroid shifts. Since the clusters are all shifted from the true center of the maps, the final cluster stack will have a lower dipole amplitude since the lensing signal got smoothed out to some extent, which leads to a bias towards lower masses. Specifically, we find a bias of $\bias = -0.05\pm0.01$ for $\sigma_\text{offset} = 0.5^{\prime}$, which is a typical offset between the SZ centers in CMB data and the red brightest cluster galaxy in  optical and near-infrared data \citep{song12}. Having an estimate of the expected centroid shift for the considered galaxy cluster survey, $\sigma_\text{expected}$, the bias can be mitigated by shifting the lensed model simulations according to a Gaussian distribution $N(0, \sigma_\text{expected})$. 
While this increases the mass uncertainties, we find that it is $\lesssim 10 \%$ for the cases $\sigma_\text{offset} \lesssim 0.5^{\prime}$, and thus negligible (see Fig. \ref{fig:cluster_positions}). 
For the typical case of $\sigma_\text{offset} = 0.5^{\prime}$, we find a relative bias value $\bias = -0.01\pm0.01$ after correction.

\subsubsection{Foreground Bias}
\label{sec:foreground_bias}
While cluster-correlated signals are largely unpolarized and were not a concern for the results in \citetalias{raghunathan19c}, they can have a large impact on the lensing mass reconstruction when using CMB temperature data. In this section, we analyze the effects of the cluster-correlated kSZ and tSZ signals on the final mass result both separately and together. For this, we add $(60'\times60')$ kSZ and/or $(60'\times60')$ 150 GHz tSZ maps, obtained from \mdplsimname{} simulations \citep{omori2022}, to the cluster-lensed maps. 
Since we are using $25,000$ lensed cutouts in this analysis, we pick clusters in the mass range $M_{200c} \in [1.25, 1.69]\times10^{14}\text{ }M_\odot$ and redshift range $z\in[0.6, 0.8]$ to get a large enough sample of kSZ and tSZ simulations. 
For this section, we set the cluster mass to $M_{200c} =1.47\times10^{14}\text{ }M_\odot$ and use a redshift $z = 0.7$ for the cluster-lensed maps.
As in the previous section, we add a white noise level $\Delta T_\text{white} = 2$ $\mu$K-arcmin to all the maps.  

Since the clusters can either have a positive or negative kSZ signal depending on their peculiar velocity relative to us, the estimator is naturally immune to kSZ-induced lensing biases when we stack the lensing signal from multiple clusters. However, including the kSZ signal slightly increases the variance in the final lensing dipole stack.
Specifically, the mass uncertainties are increased by a factor of 1.2 compared to the baseline case.

%\sout{When accounting for the tSZ effect, we do not get any reliable results, i.e. $p$-values $\simeq$ 0, since the final stack is dominated by the mean tSZ signal of all the clusters. After applying the tSZ mitigation step (see Eq. (\ref{eq:stack_tsz}) and (\ref{eq:stack_dipole_tsz})), we get $p$-values similar to the baseline and kSZ case. We find a relative bias $\bias = -0.01 \pm  0.02$, indicating that the tSZ contamination has been effectively reduced, and an increase in the mass uncertainty by a factor of $\sim 2$ compared to the baseline case. We find similar results when considering the impact of both the tSZ and kSZ signal after applying the tSZ mitigation step.}
On the other hand, the tSZ signal introduces significant issues. 
For example, when we compare the model dipole stack to the mock data stack in the presence of the tSZ signal, we get a Probability-To-Exceed (PTE) value $\simeq 0$, since the final stack is dominated by the mean tSZ signal of all the clusters. 
This improves and matches the baseline and kSZ cases
after applying the tSZ mitigation step (see Eq. (\ref{eq:stack_tsz}) and (\ref{eq:stack_dipole_tsz})).
In that case, we find a relative bias $\bias = -0.01 \pm  0.02$, indicating that the tSZ contamination has been effectively reduced. 
However, the tSZ mitigation step increases the mass uncertainty by a factor of $\sim 2$ compared to the baseline case. 
As expected, we find similar results when considering the impact of both, the kSZ and tSZ signals, after applying the tSZ mitigation step. \refresponse{Considering both the impact due to uncertainties in the cluster positions (using $\sigma_\text{offset} = 0.5^{\prime}$), as well as the kSZ and tSZ signals after applying the tSZ mitigation step, we find the mass uncertainty to increase by $\sim 2.5$ compared to the baseline case.}

Similar to \S\ref{sec:validation}, we note that the gradient direction estimation could be contaminated by the tSZ signal for massive clusters. 
\addcomments{This can be mitigated using the techniques mentioned in \S \ref{sec:validation}, or by using a tSZ-nulled map for the gradient estimation.}
Since the SNR of the CMB for modes $\ell < 2000$ is extremely high for current and future CMB surveys, the noise enhancement due to tSZ nulling will have negligible impact on the final lensing SNR. 

%It should be noted that we observe an overall shift \srini{how much?} towards lower masses after applying the tSZ mitigation step (see Fig. \ref{fig:foreground_bias}). 
%\sout{This shift is due to the residual tSZ signal in the Wiener-filtered, lensed simulations, leading to sub-optimal stacking.} This is due to the gradient contamination due to the tSZ signal in the Wiener-filtered lensed simulations which lead to differences in stacking between the mock data and model. Similar to \S\ref{sec:validation}, when we redo the analysis using the gradient orientations from the unlensed simulations, we find no shift in the mass estimates after mitigating the tSZ signal. While the relative biases after applying the tSZ mitigation step are within the statistical uncertainties for the considered case, and the tSZ contamination in CMB temperature maps is usually reduced using an ILC technique, this low-mass bias could be larger for higher cluster masses since the tSZ signal depends on the mass of the cluster. As was discussed in \S\ref{sec:validation}, inpainting can provide one possibility to get rid of this residual bias. \srini{Going to rewrite this paragraph a little bit after consulting Kevin. Also, note that we can estimate gradient from the tSZ-nulled map as Kaustav pointed out.}

\begin{figure}
    \centering
    \includegraphics[width=10cm , height=8cm]{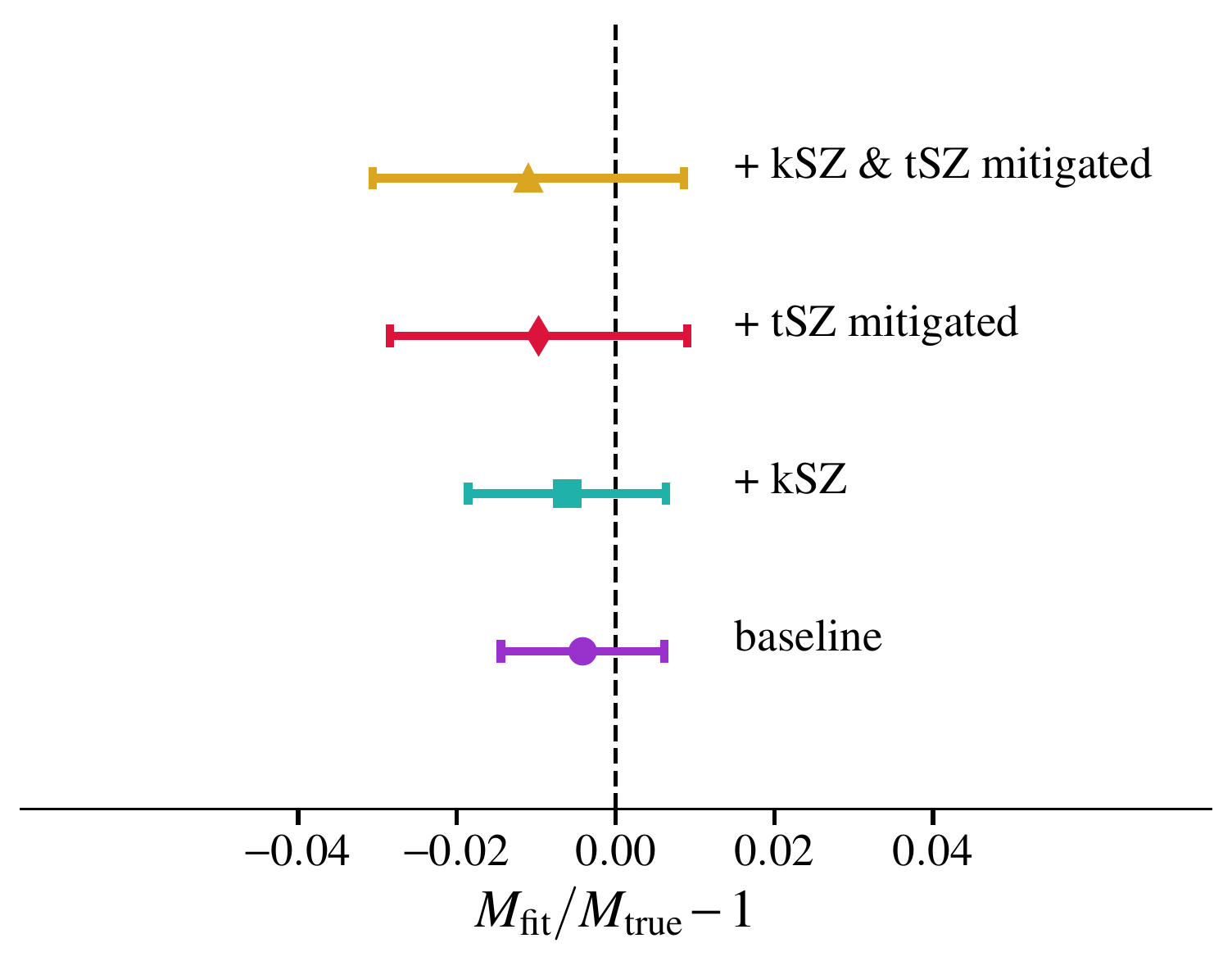}
    \caption{
    %\sout{Bias analysis due to the clusters' tSZ and kSZ signals}
    Bias analysis due to cluster-correlated kSZ and tSZ signals. The underlying CMB realizations have been fixed for all the cases to highlight the impact of the foregrounds on the mass estimates. The kSZ signal (turquoise square) only acts as an additional source of variance, while the tSZ signal has to be mitigated to get an unbiased mass estimate (red diamond).  
    }
\label{fig:foreground_bias}
\end{figure}

%%%%%%%%%%%%%%%%%%%%%%%%%%%%%%%%%%%%%%%%%%%%%%%%
%%%%%%%%%%%%%%%%%%%%%%%%%%%%%%%%%%%%%%%%%%%%%%%%
%%%%%%%%%%%%%%%%%%%%%%%%%%%%%%%%%%%%%%%%%%%%%%%%

\section{Lensing SNR Forecasts}
\label{sec:forecasts}
The next generation CMB experiments are expected to improve the CMB-cluster lensing-based mass calibration substantially \citep{louis17, raghunathan17a, madhavacheril17}. 
In this final section, we forecast the lensing SNR for \spt{} and three upcoming experiments. 

%\sout{The mock data for each experiment includes a Gaussian realization of the expected ILC residual noise power spectrum according to Eq. (\ref{eq:residual}), using the frequency bands and noise level values given in Table \ref{tab_exp_specs} and \ref{tab_atm_noise_specs}, and the corresponding weighted spectra for RGs, the CIB, diffuse tSZ and kSZ signals, and tSZ-CIB correlations, taken from} \citet{george15, reichardt21}. 
%\sout{We add the same cluster-correlated kSZ and tSZ maps to the lensed simulations which were used in \S \ref{sec:foreground_bias}, with the difference that the tSZ maps for each experiment have been weighted according to Eq. (\ref{eq:ilc_eq}) to get a realistic estimate of the cluster's remaining tSZ signal for these experiments.}
The mock data for each experiment includes the CMB lensing signal along with residual noise and foregrounds, calculated using Eq. (\ref{eq:residual}) and the frequency bands and noise levels given in Table \ref{tab_exp_specs} and \ref{tab_atm_noise_specs}. 
To make the forecasts realistic, we also include the cluster-correlated kSZ and tSZ signals. 
Note that, compared to the foreground signals in \S \ref{sec:foreground_bias}, the tSZ signals will be slightly down-weighted here because of the frequency dependent ILC weights given by Eq. (\ref{eq:ilc_weights}).

\subsection{Experiments}
\label{sec:experiments}
The CMB experiments considered in this work include

\begin{itemize}
    \item \textbf{\spt{}}: The South Pole 
Telescope\footnote{\url{https://pole.uchicago.edu/public/Home.html}} (SPT) \citep{carlstrom11} is a 10-meter telescope located at the Amundsen–Scott South Pole Station in Antarctica. \spt{} \citep{benson14, bender18, sobrin22} is the third generation receiver operating on SPT, dedicated to high-resolution observations of the CMB. The receiver contains $\sim$16,000 polarization-sensitive detectors, providing arcminute-scale resolution maps of the CMB using 95 GHz, 150 GHz, and 220 GHz frequency band centers. \spt{} was installed in 2017 and started a 6-year 1500 deg$^2$ survey in February 2018. The expected number of clusters is 7,000 \citep{raghunathan22b}. 
\item \textbf{\so{}}: 
The Simons Observatory\footnote{\url{https://simonsobservatory.org/}} (\so{}) \citep{so19} is a next-generation CMB observatory located in the Atacama Desert in Northern Chile inside the Chajnator Science Preserve at 5,200 meters. It consists of one 6-meter diameter large-aperture telescope (SO-LAT) measuring CMB temperature and polarization. %\srini{\sout{, and three 0.5-meter diameter small-aperture telescopes (SO-SAT), which will only measure CMB polarization and are thus not of interest for this work}}. 
%In late 2022 \srini{2024 may be, who knows?}, \so{} will begin a 5-year survey. 
The observed fraction of the sky will be 40\% for the LAT instrument. %\srini{\sout{and 10\% for the SAT instruments}}. 
The frequency band centers used for all the instruments are 27 GHz, 39 GHz, 93 GHz, 145 GHz, 225 GHz and 280 GHz, and the expected number of cluster is 25,000 \citep{madhavacheril17, so19, raghunathan22b}.
\item \textbf{\fyst{}}: 
The Fred Young Submillimeter Telescope\footnote{\url{https://www.ccatobservatory.org/}} (\fyst{})\citep{choi20, aravena23}, previously known as CCAT-prime, is a 6-meter diameter telescope located at 5,600 meters on the Cerro Chajnantor mountain in the Atacama Desert of Northern Chile. FYST will cover 35\%\footnote{In this work, we assume that FYST covers the exact region of SO.} of the sky area using 220 GHz, 280 GHz, 350 GHz, and 410 GHz frequency bands. 
Combining its submillimeter imaging with the millimeter imaging of SO will allow precise separation of foreground dust emission from the CMB signal. 
\item \textbf{\cmbsfourwide{}}: The fourth-generation ground-based CMB experiment\footnote{\url{https://cmb-s4.org/}} (\cmbsfourwide) \citep{cmbs4-sb1, abazajian19} will %\srini{\sout{consist of large- and small-aperture telescopes}} 
be operating at the South Pole and in the Chajnantor Plateau in the Atacama desert in Northern Chile . The %\srini{\sout{constructions will presumably start in 2023, with an}} 
anticipated year for the start of the survey is 2029. The South Pole telescopes will conduct an ultra-deep survey (S4-Ultra deep), covering 3\% of the sky, while the Atacama telescopes will conduct a wide (S4-Wide) and a deep (S4-Deep) survey with a 65\% sky coverage. For this work, we will only consider \cmbsfourwide{}, which will use 27 GHz, 39 GHz, 93 GHz, 145 GHz, 225 GHz and 280 GHz frequency band centers. The expected number of cluster is 100,000 \citep{louis17, madhavacheril17, raghunathan22, raghunathan22b}.
\end{itemize}
%The produced data will allow to constrain inflationary theories, to measure the sum of the neutrino masses, and to constrain the nature of dark energy.
%Among the scientific goals are the search for B-mode polarization pattern of CMB photons as evidence for inflation, constraining dark energy and dark matter, measuring neutrino masses, and understanding galaxy cluster evolution.
%It will operate at submillimeter to millimeter wavelengths and measure the kSZ effect of galaxy clusters, help to improve our understanding of the cosmic origins of stars, and probe multiple spectral line tracers of the interstellar medium (ISM). Additionally, c
%The project will be designed to search for primordial gravitational waves as a test of the theory of inflation, constrain dark energy and dark matter, determine neutrinos masses, and explore the time-variable millimeter-wave sky.
Table \ref{tab_exp_specs} lists the instrumental beams and the detector noise levels $\Delta T_\text{white}$ of each frequency band for the four experiments. The parameters governing the atmospheric noise ($\ell_\text{knee}$, $\alpha_\text{knee}$, and $\Delta T_\text{red}$) are listed in Table \ref{tab_atm_noise_specs}. While the four high-frequency (HF) channels of \fyst{} cannot be used independently for CMB science, they can be combined with the \so{} frequency channels to reduce the residual noise in the final maps.

\newcommand{\inspecstablecaptiontext}{Instrumental beam and detector noise specifications for the current and future experiments considered in this work.}

\ifdefined\mentiontablenotes
{
    \begin{table}
    \centering
    \caption{\inspecstablecaptiontext}
    \vspace*{2mm}
    \begin{threeparttable}
    \footnotesize{
    \resizebox{0.98\textwidth}{!}{
    \begin{tabular}{| C{2.5cm}||C{1.2cm}|C{1cm}|C{1.2cm}|C{1.3cm}||C{1.2cm}|C{1cm}|C{1.2cm}| C{1.3cm}|}
    %\begin{tabular}{| l | c | c | c | c | c | c| c | c |}
    \hline
    \multirow{3}{*}{Frequency [GHz]} & \multicolumn{4}{c||}{Beam [arcminutes]} & \multicolumn{4}{c|}{$\Delta T_{\rm white}$ [\ukam]}\\
    \cline{2-9}
    & \spt & \so & \sofyst & \cmbsfourwide & \spt & \so & \sofyst & \cmbsfourwide\\
    \hline
    
    27 & \multirow{2}{*}{-} & \multicolumn{3}{c||}{7.4} & \multirow{2}{*}{-}  & \multicolumn{2}{c|}{52.1} & 21.5 \\\cline{1-1}\cline{3-5}\cline{7-9}
    
    39 & & \multicolumn{3}{c||}{5.1} & & \multicolumn{2}{c|}{27.1} & 11.9 \\\hline%\cline{1-1}\cline{3-5}\cline{7-9}
    
    93 & - & \multicolumn{3}{c||}{2.2} & - & \multicolumn{2}{c|}{5.8} & 1.9 \\\hline
    
    95 & 1.7 & \multicolumn{3}{c||}{-} & 3.0 & \multicolumn{3}{c|}{-} \\\hline

    145 & - & \multicolumn{3}{c||}{1.4} & - & \multicolumn{2}{c|}{6.5} & 2.1 \\\hline

    150 & 1.2 & \multicolumn{3}{c||}{-} & 2.2 & \multicolumn{3}{c|}{-}
    \\\hline

    220 & 1.0 & - & 0.95 & - & 8.8 & - & 14.6 & - \\\hline
    
    %225 & \multicolumn{2}{c|}{1.0} & 0.95 & 1.0 & 8.8 & 15.0 & 14.6 & 6.9 \\\hline
    225 & - & \multicolumn{3}{c||}{1.0} & - & \multicolumn{2}{c|}{15.0} & 6.9 \\\hline
    
    280\tnote{$\ast$} & \multirow{3}{*}{-} & 0.9 & 0.75 & 0.7 & \multirow{3}{*}{-} & 37.0 & 27.5 & 16.8 \\\cline{1-1}\cline{3-5}\cline{7-9}
    
    350 & & \multirow{2}{*}{-} & 0.58 & \multirow{2}{*}{-} & & \multirow{2}{*}{-} & 104.8 & \multirow{2}{*}{-} \\\cline{1-1}\cline{4-4}\cline{8-8}
    
    410 & & & 0.50 & & & & 376.6 & \\\hline

    \end{tabular}
    }
    }
    \begin{tablenotes}
        \item[$\ast$] For the 280 GHz band that overlaps for \so{} and \fyst{}, we combine the noise power spectra using inverse \\ variance weighting.
    \end{tablenotes}
    \end{threeparttable}
    \label{tab_exp_specs}
    \end{table}
}
\else
{
	\begin{table}
	\centering
	\caption{\inspecstablecaptiontext}
	\vspace*{2mm}
	\footnotesize{
	\resizebox{0.98\textwidth}{!}{
	\begin{tabular}{| C{2.5cm}||C{1.2cm}|C{1cm}|C{1.2cm}|C{1.3cm}||C{1.2cm}|C{1cm}|C{1.2cm}| C{1.3cm}|}
	%\begin{tabular}{| l | c | c | c | c | c | c| c | c |}
	\hline
	\multirow{3}{*}{Frequency [GHz]} & \multicolumn{4}{c||}{Beam [arcminutes]} & \multicolumn{4}{c|}{$\DeltaT_{\rm white}$ [\ukam]}\\
	\cline{2-9}
    & \spt & \so & \sofyst & \cmbsfourwide & \spt & \so & \sofyst & \cmbsfourwide\\
    \hline

	27 & \multirow{2}{*}{-} & \multicolumn{3}{c||}{7.4} & \multirow{2}{*}{-}  & \multicolumn{2}{c|}{52.1} & 21.5 \\\cline{1-1}\cline{3-5}\cline{7-9}

	39 & & \multicolumn{3}{c||}{5.1} & & \multicolumn{2}{c|}{27.1} & 11.9 \\\hline%\cline{1-1}\cline{3-5}\cline{7-9}

	93 & 1.7 & \multicolumn{3}{c||}{2.2} & 3.0 & \multicolumn{2}{c|}{5.8} & 1.9 \\\hline

	145 & 1.2 & \multicolumn{3}{c||}{1.4} & 2.2 & \multicolumn{2}{c|}{6.5} & 2.1 \\\hline

    225 & 1.0 & 1.0 & 0.95 & 1.0 & 8.8 & 15.0 & 14.6 & 6.9 \\\hline
        
    280 & \multirow{4}{*}{-} & 0.9 & 0.75 & 0.7 & \multirow{3}{*}{-} & 37.0 & 27.5 & 16.8 \\\cline{1-1}\cline{3-5}\cline{7-9}
    
	350 & & \multirow{2}{*}{-} & 0.58 & \multirow{2}{*}{-} & & \multirow{2}{*}{-} & 104.8 & \multirow{2}{*}{-} \\\cline{1-1}\cline{4-4}\cline{8-8}

	410 & & & 0.50 & & & & 376.6 & \\\hline

	\end{tabular}
	}
	}
	\label{tab_exp_specs}
	\end{table}
}
\fi

\ifdefined\mentiontablenotes
{
\begin{table}
\centering
\caption{Atmospheric noise specifications for the current and future experiments considered in this work.}
\vspace*{2mm}
\footnotesize{
%\resizebox{0.98\textwidth}{!}{
\begin{threeparttable}
\begin{tabular}{| C{1.5cm}||C{1.2cm}|C{1cm}|C{1.3cm}|C{1.2cm}|C{1cm}|C{1.3cm}||C{1cm}|C{1.8cm}|}
\hline
Frequency & \multicolumn{3}{c|}{$\ell_{\rm knee}$} & \multicolumn{3}{c||}{$\alpha_{\rm knee}$} & \multicolumn{2}{c|}{$\Delta T_{\rm red}$ [\ukam]}\\
\cline{2-9}
[GHz] & \spt & SO & \cmbsfourwide & \spt & SO & \cmbsfourwide & \so & \sofyst \\\hline

27 & \multirow{2}{*}{-} & \multirow{6}{*}{1000} & 415 & \multirow{3}{*}{-} & \multirow{6}{*}{-3.5} & \multirow{6}{*}{-3.5} & \multicolumn{2}{c|}{6.1} \\\cline{1-1}\cline{4-4}\cline{8-9}

39 & & & 391 & & & & \multicolumn{2}{c|}{3.8} \\\cline{1-2}\cline{4-5}\cline{8-9}

93\tnote{$\ast$} & 1200 & & 1932 & -3 & & &  \multicolumn{2}{c|}{9.3}\\\cline{1-2}\cline{4-5}\cline{8-9}

145\tnote{$\ast$} & 2200 & & 3917 & -4 & & &  \multicolumn{2}{c|}{23.8}\\\cline{1-2}\cline{4-5}\cline{8-9}

225\tnote{$\ast$} & 2300 & & 6740 & -4 & & & 80.0 & 434.8 \\\cline{1-2}\cline{4-5}\cline{8-9}

280 & \multirow{3}{*}{-} & & 6792 & \multirow{3}{*}{-} & & & 108.0 & 1140.2 \\\cline{1-1}\cline{3-4}\cline{6-6}\cline{8-9}

350 & & \multirow{2}{*}{-} & & & \multirow{2}{*}{-} & & \multirow{2}{*}{-} & 5648.8 \\\cline{1-1}\cline{7-7}\cline{9-9}

410 & & & & & & & & 14174.2 \\\hline

\end{tabular}
    \begin{tablenotes}
        \item[$\ast$] For simplicity, we do not include all the frequency bands in this table by assuming 93 $\leftrightarrow$ 95 GHz, 145 $\leftrightarrow$ 150 GHz, and 220 $\leftrightarrow$ 225 GHz. 
    \end{tablenotes}
\end{threeparttable}
%}
}
\label{tab_atm_noise_specs}
\end{table}
}
\fi

\subsection{Results}
\label{sec:lensing_snr}
\begin{figure}
    \centering
    \includegraphics[width=12cm , height=8.cm]{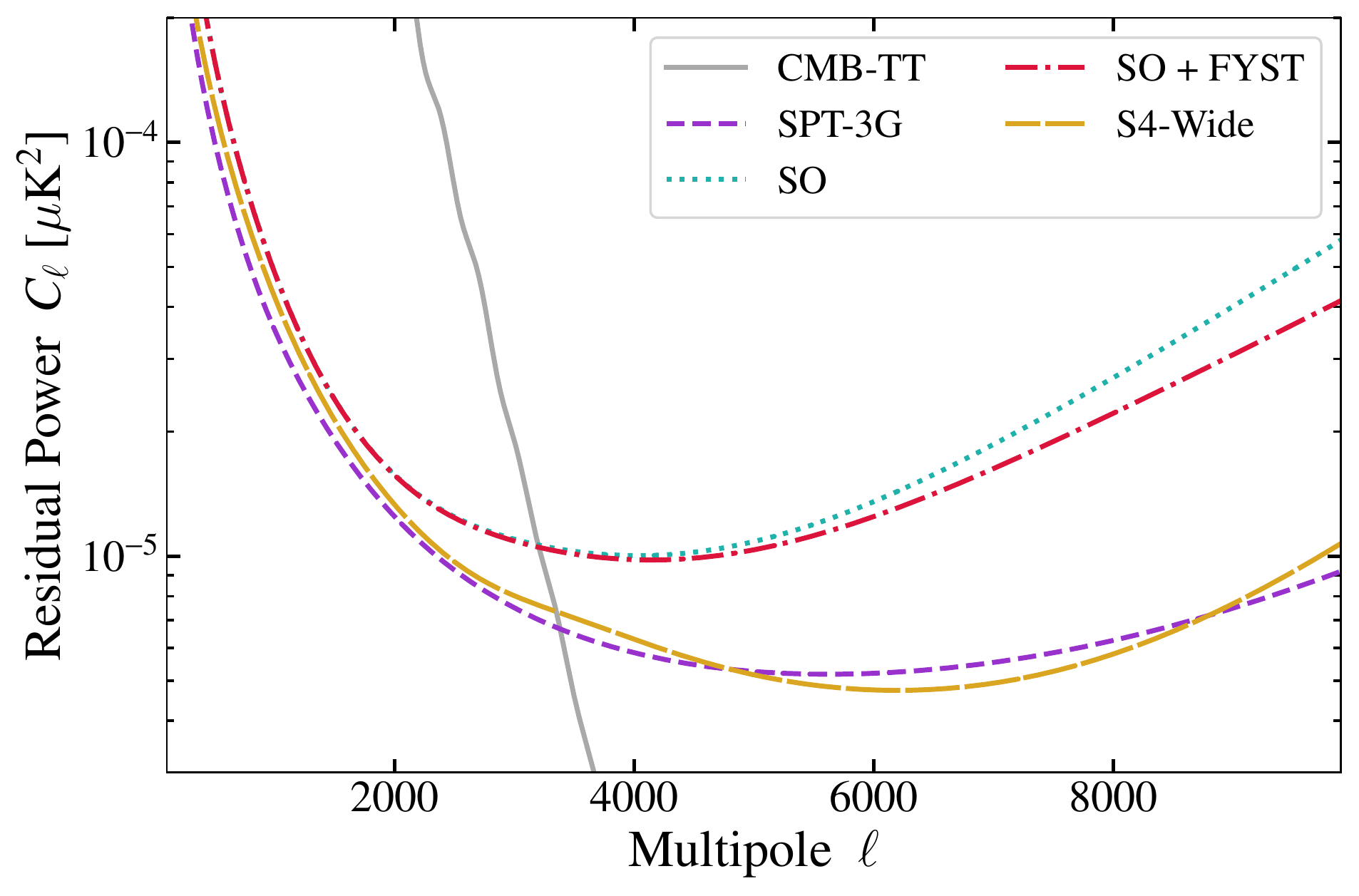}
    \caption{ILC noise power spectra for the CMB experiments considered in this work. Also shown is the CMB temperature power spectrum (solid gray line).
    \spt{} (dashed purple line) and \cmbsfourwide{} (long-dashed yellow line) are roughly similar since they are dominated by the residual CIB signals which set a noise floor. 
    The residual power for \so{} (dotted turquoise line) can be slightly decreased using the additional HF channels from FYST (dash-dotted red line).}
\label{fig:residuals}
\end{figure}

As in the previous section, we use $N_\text{rand} = 50,000$ random cutouts for the background stack. For the lensed stacks, we use the expected cluster number for each experiment: $N_{\text{clus}}^{\rm \spt} = 7,000$, $N_{\text{clus}}^{\rm \so}$ = $N_{\text{clus}}^{\rm \sofyst} = 25,000$, and $N_{\text{clus}}^{\rm \cmbsfourwide} = 100,000$. We use a cluster mass $M = 1.47\times10^{14} \text{ }M_\odot$ and a cluster redshift $z = 0.7$, since this is the mean mass and redshift used to extract the kSZ and tSZ signals from \mdplsimname{} simulations. After adding a Gaussian realization of the expected ILC residual noise power spectrum (see Fig. \ref{fig:residuals}), and the kSZ and ILC weighted tSZ maps to the cluster-lensed simulations, we smooth each map by a Gaussian instrumental beam with a FWHM corresponding to the beam at the 145 GHz or 150 GHz channel of the experiments given in Table \ref{tab_exp_specs}. As was done in \S \ref{sec:foreground_bias}, we get an estimate of the mean tSZ signal using Eq. (\ref{eq:stack_tsz}) and compute the lensing dipole according to Eq. (\ref{eq:stack_dipole_tsz}). 

\begin{figure}
    \centering
    \includegraphics[width=12cm , height=8cm]{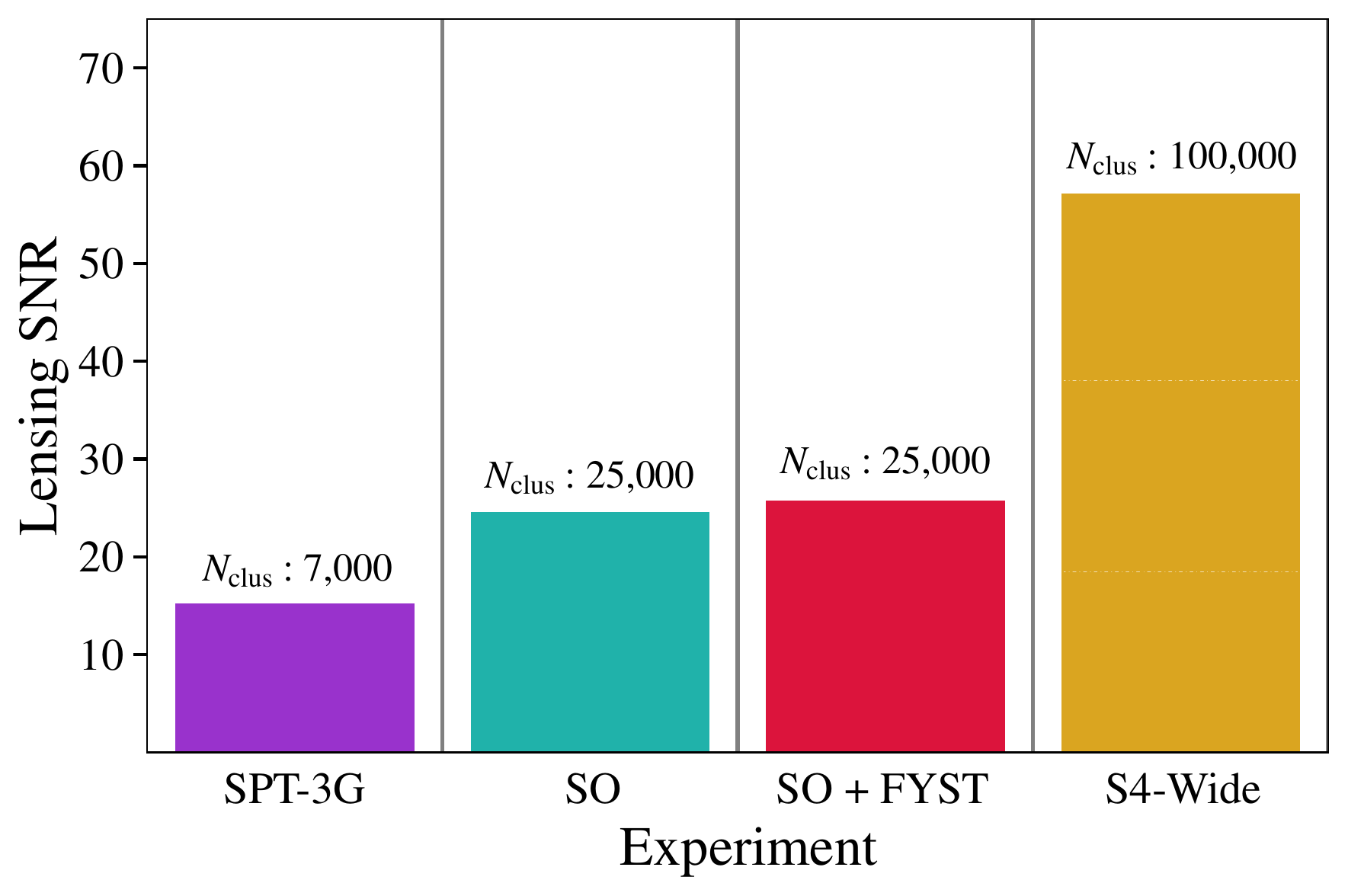}
    \caption{Expected CMB-cluster lensing SNR values for current (\spt) and future experiments (\so{}, \sofyst{}, \cmbsfourwide). 
    The SNR values for \so{} and \sofyst{} are nearly twice as high as the one for \spt. 
    For \cmbsfourwide, the sensitivity is $\times4$ higher compared to \spt.
    However, this increase in the lensing SNR for future experiments is driven by the much larger sky coverage providing more clusters to stack. 
    We note that the sensitivity for a single cluster is roughly the same for both \spt{} and \cmbsfourwide.}
\label{fig:snr}
\end{figure}

The lensing SNR for each experiment is obtained as 
\begin{align}
    \text{SNR} = \sqrt{\Delta \chi^2} = \sqrt{2\left[{\mathrm \ln} \mathcal{L}(M=M_\text{fit})-{\mathrm \ln} \mathcal{L}(M = 0)\right]} \;.
    \label{eq:snr}
\end{align}
Specifically, we find SNR$_{\rm \spt}$ = 15.1 (6.6\% mass constraints), SNR$_{\rm \so}$ = 24.4 (4.1\% mass constraints), SNR$_{\rm \sofyst}$ = 25.6 (3.9\% mass constraints), and SNR$_{\rm \cmbsfourwide}$ = 57.0 (1.8\% mass constraints). The resulting SNR values are shown in Fig. \ref{fig:snr}. 
%\sout{Combining the \so{} and \fyst{} frequency channels only slightly increases the lensing SNR.} 
Adding the HF information from \fyst{} to \so{} has little impact on the lensing SNR. 
However, we note that the HF channels can give a better handle on the foregrounds for cluster detection.
%\sout{Therefore, the frequency channels of \so{} alone should be sufficient for CMB-cluster lensing measurements.} 
It is expected that \so{} will detect nearly four times more clusters than \spt{} due to a much larger sky coverage, resulting in a SNR value roughly twice the value of \spt{}. On the other hand, S4-Wide will have a $\times14$ larger cluster sample than \spt{}, which leads to a significantly higher lensing SNR (roughly $\times4$ higher than for \spt{}). 
%\sout{It should be noted that the cluster sample size for each experiment is not restricted to the SZ selected clusters of the experiments, as the samples can be increased by selecting clusters from optical and X-ray surveys to further increase the lensing SNR.}
%\srini{I think this last sentence is not required here. May be in conclusion / discussion.}

%%%%%%%%%%%%%%%%%%%%%%%%%%%%%%%%%%%%%%%%%%%%%%%%
%%%%%%%%%%%%%%%%%%%%%%%%%%%%%%%%%%%%%%%%%%%%%%%%
%%%%%%%%%%%%%%%%%%%%%%%%%%%%%%%%%%%%%%%%%%%%%%%%

\section{Conclusions}
\label{sec:conclusion}

In this work, we extended the cluster lensing estimator developed in \citetalias{raghunathan19c} to CMB temperature data. 
We compared the performance of the estimator to the MLE and QE, finding that the SNR of the current estimator matches the MLE for all map noise levels. 
We also analyzed the impact of the cluster-correlated kSZ and tSZ signals.
Unlike other CMB-cluster lensing estimators, we showed that the kSZ signal does not introduce a bias and only acts as an additional source of variance. 
We also demonstrated that the bias induced by the tSZ signal can be trivially mitigated by subtracting a tSZ estimate obtained from the cluster-lensed cutouts themselves. \addcomments{When working with real data, this technique will also help in removing the bias due to the CIB, as well as any other foregrounds correlated with the clusters under study.}

We did not include the effects of the rotational kSZ \citep{chluba02, baldi18, baxter19, altamura23} or the moving-lens effect \citep{hotinli19, yasini19}, which are also expected to produce a dipole signal. 
However, the orientation of the dipole will be aligned along the direction of the cluster rotation in the former and the direction of the large-scale velocity field in the latter, which are both not correlated with the direction of the background CMB gradient. 
As a result, these signals tend to dilute in our final stack.
Moreover, the amplitude of these signals are expected to be similar in level or smaller than cluster lensing for the typical clusters considered in this work. Therefore, the level of bias should be negligible compared to the statistical errors.

We also studied the impact of mis-centering and found a bias $\bias = -0.05\pm0.01$ for realistic positional uncertainties in the cluster location ($\sigma_\text{offset} = 0.5^{\prime}$). 
However, this bias can be easily mitigated by including the positional uncertainties in the model lensing dipoles with a negligible degradation of the lensing SNR for $\sigma_\text{offset} \lesssim 0.5^{\prime}$.

Finally, we presented forecasts for the expected lensing SNR for current and future CMB surveys. We predict cluster mass uncertainties of 6.6\% for \spt{} with 7,000 clusters, 4.1\% for \so{} and 3.9\% for \so{} + \fyst{} with 25,000 clusters, and 1.8\% for \cmbsfourwide{} with 100,000 clusters. 
The additional HF bands from \fyst{} did not significantly increase the lensing SNR of \so{}. However, they might be useful to handle the contamination of the tSZ signal due to dusty galaxies for cluster detection. 

The lensing estimator presented in this work can also be applied for mass calibration of clusters identified from optical or X-ray surveys. 
%\addcomments{Maybe mention that this estimator will be applied on real CMB temperature data?}

%%%%%%%%%%%%%%%%%%%%%%%%%%%%%%%%%%%%%%%%%%%%%%%%
%%%%%%%%%%%%%%%%%%%%%%%%%%%%%%%%%%%%%%%%%%%%%%%%
%%%%%%%%%%%%%%%%%%%%%%%%%%%%%%%%%%%%%%%%%%%%%%%%

\section*{Acknowledgements}
\label{sec:ack}

We thank Behzad Ansarinejad and Christian Reichardt for useful discussions. 
We thank Yuuki Omori for providing access to \mdplsimname{} simulations \citep{omori2022}.
\addcomments{We also thank the anonymous referee for their feedback and suggestions that helped in improving the flow of this manuscript.}

This work was partially supported by the Center for AstroPhysical Surveys (CAPS) at the National Center for Supercomputing Applications (NCSA), University of Illinois Urbana-Champaign. 
This work made use of the following computing resources: Illinois Campus Cluster, a computing resource that is operated by the Illinois Campus Cluster Program (ICCP) in conjunction with the National Center for Supercomputing Applications (NCSA) and which is supported by funds from the University of Illinois at Urbana-Champaign; the computational and storage services associated with the Hoffman2 Shared Cluster provided by UCLA Institute for Digital Research and Education's Research Technology Group; and the University of Melbourne’s Research Computing Services and the Petascale Campus Initiative.

%%%%%%%%%%%%%%%%%%%%%%%%%%%%%%%%%%%%%%%%%%%%%%%%
%%%%%%%%%%%%%%%%%%%%%%%%%%%%%%%%%%%%%%%%%%%%%%%%
%%%%%%%%%%%%%%%%%%%%%%%%%%%%%%%%%%%%%%%%%%%%%%%%

\bibliography{references}

\end{document}